\documentclass[aps,superscriptaddress,floatfix]{revtex4}

\usepackage{graphicx}
\usepackage{epsfig}
\usepackage{amssymb}

\begin{document}
\newcommand{\be}{\begin{eqnarray}}
\newcommand{\ee}{\end{eqnarray}}
\newcommand\del{\partial}
\newcommand\nn{\nonumber}
\newcommand{\Tr}{{\rm Tr}}
\newcommand{\mat}{\left ( \begin{array}{cc}}
\newcommand{\emat}{\end{array} \right )}
\newcommand{\vect}{\left ( \begin{array}{c}}
\newcommand{\evect}{\end{array} \right )}
\newcommand{\tr}{\rm Tr}
\def\conj#1{{{#1}^{*}}}
\newcommand\hatmu{\hat{\mu}}

\voffset 2cm

\title{The QCD Sign Problem for Small Chemical Potential}

\author{K. Splittorff}
\affiliation{The Niels Bohr Institute, Blegdamsvej 17, DK-2100, Copenhagen {\O}, Denmark}
\author{J.J.M. Verbaarschot$^{1,\, }$}
\affiliation{Niels Bohr International Academy, Blegdamsvej 17, DK-2100, Copenhagen {\O}, Denmark}
\affiliation{Department of Physics and Astronomy, SUNY, Stony Brook,
 New York 11794, USA}

\date   {\today}
\begin  {abstract}
The expectation value of the complex phase factor of the fermion determinant 
is computed in the microscopic domain of QCD at nonzero chemical potential.
We find that the average phase factor is non-vanishing below a critical
value of the chemical potential equal to the half the pion mass and
vanishes exponentially in the volume 
for larger values of the chemical potential.
This holds for QCD with dynamical quarks as well 
as for  quenched and phase quenched QCD. 
The average phase factor has an essential singularity 
 for zero chemical potential and cannot be obtained by analytic continuation
from imaginary chemical potential or by means of a Taylor expansion.
The leading order correction in the $p$-expansion of the chiral
Lagrangian is calculated as well.
\end{abstract}

\maketitle
\newpage
 
\section{Introduction}

A nonzero baryon density in QCD is achieved by introducing a chemical
potential which enhances the propagation of quarks in the forward time 
direction and inhibits their propagation in the backward time direction. 
This imbalance makes the fermion determinant in the Euclidean formulation 
of QCD complex. 
The integration measure in the QCD partition function therefore cannot 
be used directly to define a probabilistic measure to generate gauge field 
configurations. Because of this {\sl sign problem}, standard lattice QCD 
does not apply at nonzero baryon chemical potential. This does not mean 
that lattice
QCD simulations at nonzero chemical potential are all altogether impossible.
If the complex phase factor of the fermion determinant does not 
fluctuate strongly the sign problem may become manageable. 
The reweighting method \cite{Glasgow} deals with the sign problem 
by including the phase factor in the observable rather than in the measure.  
While this method works in principle, its limitations 
are set by the strength of the fluctuations of the phase of the fermion
determinant.
It is therefore of considerable interest to understand how the fluctuations
of the phase factor depend on the temperature, volume,
chemical potential, quark mass as well as the lattice cutoff.     

In this paper we study the severity of the sign problem in QCD as
expressed through the vacuum expectation value of the phase factor
\be
e^{2i\theta}=\frac{\det(D +\mu\gamma_0 +m)}{\det(D^\dagger +\mu\gamma_0 +m)}.
\label{phasedef}
\ee
Since the average phase factor is the ratio of two partition functions,
it is typically exponentially small in the volume requiring that 
the number of configurations needed in 
a reweighting approach is exponentially large with the volume.

We will consider low temperatures where the chiral 
condensate $\Sigma$ is nonzero
and pions dominate the excitations of the 
QCD vacuum. We will take
the square of the chemical potential 
inversely proportional to the volume
\be
 \mu^2F_\pi^2V \sim 1,
\label{mumicro}
\ee
where $F_\pi$ is the pion decay constant. If furthermore
\be
 m\Sigma V \sim 1 \quad {\rm and} \quad  V\Lambda_{\rm QCD}^4 \gg 1,
\label{mmicro}
\ee
the QCD partition function reduces to the zero momentum limit
of the corresponding chiral Lagrangian.
The scaling of $\mu$ and $m$ in (\ref{mumicro}) and (\ref{mmicro}) defines
the microscopic limit of QCD and is also referred to as the 
$\epsilon$-regime \cite{GLeps,LS}.  
In this limit, the zero momentum 
modes of the pions factorize from the QCD(-like)  partition function
and exact analytical expressions can be obtained. We will derive  exact
expressions for the average phase factor in both quenched QCD and QCD with
dynamical quarks
as well as in its phase quenched version. The result in all cases is that the
expectation value of the phase factor is exponentially small in the volume
when 
$\mu>m_\pi/2$. For chemical potentials smaller than half the pion mass the
average phase factor is nonzero. This suggests
that the exact analytical predictions for $\mu<m_\pi/2$ can be tested by
means of lattice QCD even in the presence of dynamical quarks. 

In the past few years several lattice QCD methods have been developed to 
study the properties of QCD at nonzero chemical potential. In particular, 
progress has been made with reweighting \cite{fodor1,fodor2}, imaginary 
chemical potential \cite{owe1,owe2,maria} the Taylor
expansion method \cite{Allton1,Allton2,Allton3,gupta} and the density of
states method \cite{schmidt} (for a recent review see \cite{schmidtrev}). 
Generally, one expects these methods to work if the average phase of the 
quark determinant fluctuates only mildly. For this reason several lattice 
measurements of the phase have appeared 
\cite{schmidt,Toussaint,deFL,NakamuraPhase,Ejiri1,Ejiri2,Gibbs}. In
this paper we will find that the quenched as well as the unquenched average 
of the phase factor has an essential singularity at zero chemical 
potential and cannot be 
obtained by analytic continuation from imaginary chemical potential 
or by means of a Taylor expansion. We give the explicit form of the 
non-analytic part and show that it is responsible for the abrupt 
change of the average phase factor at $\mu=m_\pi/2$.   
We also give the exact predictions for purely imaginary $\mu$. 
Since these are analytic in $\mu$ they can be tested directly with 
simulations at imaginary $\mu$ or by means of Taylor expansion. 
We hope that the predictions for the average phase both for $\mu^2>0$ 
and for $\mu^2<0$ will motivate lattice studies of the phase fluctuations
in the microscopic domain and thereby deepen our understanding of QCD at 
nonzero chemical potential.  

\vspace{2mm}

The scale $\mu=m_\pi/2$ has been part of lattice QCD at nonzero chemical
potential starting from the very first calculations \cite{early,Gibbs}. 
Although it manifests itself in
different ways in quenched and unquenched QCD, it has the same physical
origin, namely condensation of charged pions. 

The most direct effect of pion condensation occurs in QCD with an equal
number of quarks and conjugate quarks (referred to as phase quenched QCD).   
For a conjugate quark, the fermion determinant is replaced by its complex 
conjugate, and therefore this theory corresponds to ignoring the phase
factor of the fermion determinant. 
Since complex conjugation of a fermion determinant corresponds 
to changing the sign of the chemical potential
\cite{AKW}, two-flavor 
phase quenched QCD is identical to QCD with a  nonzero isospin
chemical potential.  Because the pions have a nonzero isospin charge and are
weakly interacting they will Bose condense for $\mu>m_\pi/2$
at zero temperature \cite{KST,KSTVZ,SS,eff}.

The quenched approximation at nonzero chemical potential is the zero
flavor limit of QCD with an equal number of quarks and conjugate quarks
\cite{Girko,misha}. It therefore inherits the scale $\mu=m_\pi/2$. For example, one
finds that the quark number density is nonzero for $\mu>m_\pi/2$
\cite{early,LKS}.

In QCD with dynamical quarks the Dirac spectrum for typical gauge field 
configurations has the same support as in the quenched case. Therefore,
for quark masses such that $m_\pi/2<\mu$, one or more of the quark masses
are inside the domain of the Dirac eigenvalues, and the sign 
problem is severe. It has to be like that because the phase factor has to
wipe out the Bose condensate of the phase quenched theory (see also the
review \cite{splitrev}). 
In QCD with dynamical quarks, the scale $\mu=m_\pi/2$ is thus the  scale where
 phase fluctuations become strong. This is also illustrated by the spectral
density of full QCD:
When the quark mass is inside the support of the eigenvalues the unquenched 
spectral density becomes a complex function with 
strong oscillations in a finite domain of its support \cite{O,AOSV}. 
Physically these oscillations, with an amplitude that diverges exponentially
with the volume and a period inversely proportional to the volume,
are important because they are responsible for 
the discontinuity in the chiral condensate \cite{OSV}. 

The difference between the phase quenched theory (where the 
phase of the fermion determinant is ignored) and unquenched QCD gives an 
indication of  the importance of the phase factor.
In fact the ratio of the unquenched partition function and the phase
quenched partition function for two flavors is exactly the expectation
value of the phase factor in the phase quenched theory (\ref{phasedef}). 
For $\mu < m_\pi/2$ and zero temperature the free
energy of the two partition functions is the same in the thermodynamic
limit, but for $\mu > m_\pi/2$, because of the formation of a Bose
condensate in the phase quenched theory, the two free energies become
different. This results in an exponential suppression of the average
phase factor.
In this paper we will show that the same behavior is found for the expectation
value of the phase factor with respect to the full QCD partition function
and the quenched QCD partition function. We will do this by an explicit 
calculation of the average phase factor in the microscopic domain of QCD.

\vspace{2mm}

In the microscopic limit QCD is equivalent to a chiral random matrix model
with the same global symmetries \cite{SV,V,Vplb,O}. The reason is that
the static part of the chiral Lagrangian is uniquely determined by global
symmetries.
We exploit this
equivalence to calculate the average phase factor in this domain.
Alternatively, we could have started from the static part of a chiral
Lagrangian and performed the integral over the zero momentum 
modes of the Goldstone
fields. Since the phase factor includes a bosonic determinant
 at nonzero chemical
potential this method is, however, technically  demanding and has not
been worked out up to now.    

\vspace{2mm}

In the microscopic limit it is useful to consider a fixed topological 
charge sector
rather than a fixed vacuum angle \cite{LS}. For simplicity, in this paper we
will only consider the sector of zero topological charge. The results for
arbitrary topological charge follow by an immediate generalization. 

\vspace{2mm}
Part of the results derived in this paper were announced in the
letter \cite{exp2ith-letter}. In addition to providing details of the
derivations, we obtain below explicit expressions for the average phase
factor and show that it is a non-analytic function of the chemical
potential.
We also compute the  corrections to the microscopic results  to leading
order in the $p$-expansion of the chiral Lagrangian.

\vspace{2mm}

The behavior of the average phase factor can be understood from  general
arguments presented in the next section. The connection between these 
general arguments and the exact expressions in the microscopic
limit are explained in section \ref{sec:pq} where we analyze the 
average phase factor in the phase quenched theory. In sections 
\ref{sec:rmt} and \ref{sec:exp2ith_micro} we 
derive exact expressions for the average phase factor in the microscopic
limit. The thermodynamic limit of these results is evaluated in section 
\ref{sec:exp2ith_ther}, and  the results obtained from general 
arguments will be confirmed. 
Section \ref{sec:lattice} contains a discussion of the analytic
properties of the average phase factor, and 
we finish with concluding remarks. Technical details are worked out in
two appendices.

\section{The average phase factor from a general argument}
\label{sec:gen}

In this section we will consider the phase fluctuations of the 
fermion determinant in the QCD partition function 
\be
Z_{N_f} =\left\langle {\det}^{N_f} (D+\mu\gamma_0 + m) \right\rangle,
\ee
using mean field arguments at low energy. (Throughout this manuscript
$\langle\ldots\rangle$ denotes the quenched average. Dynamical averages are
labeled by a subscript referring to the number of flavors $N_f$.) 
In order to quantify the phase fluctuations 
of the fermion determinant we compute
the average phase factor \footnote{In order to write the denominator of the
  $\langle\ldots\rangle$ as a bosonic  
integral, it has to be ``hermiticized'' \cite{Janik,Feinberg} which requires 
the introduction of an additional bosonic and fermionic determinant.
Since the contribution of the additional fermion and boson cancels trivially
in the mean field argument we can safely ignore this.} 
\be
\langle e^{2i\theta} \rangle_{N_f} = \frac 1{Z_{N_f}}
\left \langle \frac{\det(D +\mu\gamma_0 +m)}
   {\det(D^\dagger +\mu\gamma_0 +m)} {\det}^{N_f}(D + \mu\gamma_0+m)\right \rangle.
\label{phase}
\ee
Notice that the average phase factor is the ratio of two partition
functions: the partition function with an additional fermionic quark and an
additional conjugate bosonic quark and  the ordinary QCD partition
function
\be
\langle e^{2i\theta} \rangle_{N_f} = \frac{Z_{N_f+1|1^*}}{Z_{N_f}}.
\label{phase-fram-Zs}
\ee
This implies that the average phase factor is necessarily real.
We will consider the low temperature limit with quark masses and chemical 
potentials well below $\Lambda_{\rm QCD}$. Then the partition function can be
described in terms of a chiral Lagrangian.
In this section we will focus on the mean field
limit where the fields are replaced by their saddle point values. In this
limit the partition functions assumes the general form 
\be
Z \sim J\left(\prod_k \frac 1{{m_\pi(\mu)}}\right) e^{-V F},
\label{gen1}
\ee
where the Jacobian, $J$, is from the measure of the Goldstone manifold at the
saddle point, and $m_\pi(\mu)$ are the $\mu$ dependent masses of the Goldstone
modes. (For a discussion of the parameterization 
dependence of the masses of the 
Goldstone modes see Appendix \ref{app:jmass}.) 
The free energy density at the saddle point, $F$, is intensive, and the 
prefactor can be written as a $1/V$ correction to the free energy.

The QCD partition function, $Z_{N_f}$, in this approximation is
completely independent of the chemical potential; since the pions have zero
baryon charge neither the free energy density nor the exponential
prefactor depend on $\mu$. 

At zero chemical potential the fermion determinant is real so in this case
the two additional determinants in $Z_{N_f+1|1^*}$ cancel and the average
phase factor is 1. When the chemical potential is nonzero, the two additional
flavors become important. The combinations of the conjugate bosonic quark
with  any of the fermionic quarks forms a charged Goldstino. 
Because of the condensation of charged Goldstinos for $\mu > m_\pi/2$
we need to consider separately the cases 
$\mu<m_\pi/2$ and $\mu>m_\pi/2$.

For $\mu<m_\pi/2$ the free
energy $F$ is independent of  the chemical potential. In the expectation
value of the phase factor the contributions of the free energy 
to the
average phase cancel. The prefactor in (\ref{gen1})
however depends on $\mu$ because the  $\mu$-dependence of the masses
of the charged Goldstone modes.
Half of the $2(N_f+1)$
charged modes have masses $m_\pi(\mu)=m_\pi+2\mu$ while the other half have
masses $m_\pi(\mu)=m_\pi-2\mu$. The mean field result for the ratio of
$Z_{N_f+1|1^*}$ and $Z_{N_f}$ is thus given by
\be
\langle e^{2i\theta}\rangle_{N_f} =
\frac {(m_\pi -2\mu)^{N_f+1}(m_\pi +2\mu)^{N_f+1}}
{m_\pi^{2N_f+2}}= (1-\frac{4\mu^2}{m_\pi^2})^{N_f+1}
\qquad {\rm for} \qquad \mu < \frac 12 m_\pi .
\ee

For $2\mu>m_\pi$ the free energy of $Z_{N_f+1|1^*}$ depends
of $\mu$.  The free 
energy has the usual $\mu$ dependence for the static part 
of a chiral Lagrangian in
the condensed phase (see for example \cite{KST,KSTVZ,eff}). 
Subtracting the free energy at $\mu=0$ we find
\be
\langle e^{2i\theta} \rangle_{N_f} \sim e^{-2VF^2_\pi\mu^2(1
  -m^2_\pi/4\mu^2)^2}\qquad {\rm for} \qquad \mu > \frac 12 m_\pi.       
\label{exp2ith_therm_cond}
\ee
The average phase factor is thus suppressed exponentially with 
the volume for $\mu>m_\pi/2$. Care has to be taken to compute 
the exponential prefactor. For 
$\mu>m_\pi/2$ there are massless modes associated with the condensation of
pions. Consequently the leading contribution to
the prefactor given in (\ref{gen1}) vanishes. One way to compute the
subleading contributions to the prefactor is from the thermodynamic limit 
of the exact microscopic result. In the next section we perform this 
calculation in the phase quenched theory. 

In section \ref{sec:exp2ith_ther} we will confirm the results obtained here 
from the thermodynamic limit  of the exact microscopic expressions.

\section{The Average Phase Factor in the Phase Quenched Theory}
\label{sec:pq}

In this section we will discuss the chemical potential dependence of the
average phase factor in the phase quenched theory, which can be 
easily obtained from known results.
First we will discuss
the exact microscopic result and then relate it         
to the mean field results of section \ref{sec:gen}.

The phase quenched partition function with two dynamical flavors 
is just the two-flavor QCD partition function without
the phase of the fermion determinant, 
\be
Z_{1+1^*}(m;\mu) = \left\langle|\det(D+\mu\gamma_0+m)|^2\right\rangle.
\ee 
This partition function is equivalent to QCD at nonzero isospin chemical
\cite{AKW}.
The expectation value of the phase factor of the fermion determinant 
for the phase quenched theory is thus given by
\be
\langle e^{2i\theta}\rangle_{1+1^*} =  
\frac{ \langle {\det}^2(D+\mu \gamma_0+m )\rangle}
{\langle |\det(D+\mu \gamma_0 +m)|^2\rangle} = \frac {Z_{N_f=2}}{Z_{1+1^*}}.
\label{exp2ithPQ}
\ee
Its microscopic limit therefore follows immediately from the microscopic limit 
 of the two  partition functions.

The microscopic limit of the phase quenched partition function is given
by \cite{SplitVerb2}
\be
Z_{1+1^*}(\hat m;\hat \mu) = 2e^{2\hat \mu^2} 
\int_0^1 dt t e^{-2\hat\mu^2 t^2} I_0^2(\hat m t),
\label{pq-micro}
\ee
where
\be
\hat m = m V \Sigma \qquad {\rm and} \qquad \hat \mu = \mu F_\pi \sqrt V.
\label{def-micro}
\ee
The two-flavor partition function is $\mu$-independent at
scales well below the nucleon mass. Its microscopic limit is given by
\cite{LS}
\be
Z_{N_f=2}(\hat{m}) = I_0^2(\hat m) - I_1^2(\hat m).
\ee
The  $\mu$-dependence of the phase factor thus follows from the
$\mu$-dependence of the phase quenched partition 
function (\ref{pq-micro}) 
\be
\langle e^{2i\theta}\rangle_{1+1^*} 
= \frac{Z_{N_f=2}(\hat{m})}{Z_{1+1^*}(\hat{m};\hat\mu)}
= \frac{I_0^2(\hat{m})-I_1^2(\hat{m})}
{2e^{2\hat \mu^2}\int_0^1 dt t e^{-2\hat\mu^2 t^2} I_0^2(\hat m t)}. 
\label{pq-phase}
\ee
\vspace{2mm}

We now take the thermodynamic limit, $\hat{m}\to\infty$ and
$\hat\mu^2\to\infty$, of the microscopic result for the phase factor 
in the phase quenched theory. In this limit the phase quenched theory 
has a second order phase transition at $\hat m /(2\hat \mu^2) = 1$. 
To show this we calculate the integral in (\ref{pq-micro}) by a saddle 
point approximation. The transition between the two phases takes place 
when the saddle point hits the boundary of the integration region, i.e. 
when 
\be
\bar t = \frac {\hat m}{2\hat \mu^2} =1.
\ee
If we use the Gell-Mann$-$Oakes$-$Renner  
relation we find that this corresponds to the expected
critical value of the chemical potential of $\mu=m_\pi/2$.   

For $\mu<m_\pi/2$, the saddle point is outside the integration region and the
leading contribution to the integral comes from the edge of the integral at 
$t=1$.
In the thermodynamic limit we obtain (this is derived 
under the assumption that the integral is cut off by the exponential factor
rather than the Gaussian factor, which in the thermodynamic limit is violated
only very close 
to the critical point) 
\be
Z_{1+1^*}(\hat{m};\hat\mu) 
\sim \frac{1}{2\pi\hat{m}(\hat{m}-2\hat\mu^2)}e^{2\hat m}\qquad
{\rm for }\qquad \hat m > 2\hat\mu^2. 
\label{n=1norm}
\ee
We find that for $\hat m > 2\hat \mu^2$ the free energy of 
the phase quenched theory is $\mu$ independent in the thermodynamic
limit. The thermodynamic limit of the microscopic two flavor partition
function is  
\be
Z_{N_f = 2}(\hat{m};\hat\mu) \sim \frac{1}{2\pi\hat{m}^2}e^{2\hat m}.
\ee  
In the average phase (\ref{pq-phase}) the free energies  
cancel. The $\mu$-dependence only enters through $1/V$ corrections,    
i.e. through the pre-exponential factors
\be
\langle e^{2i\theta}\rangle_{1+1^*} \sim (1 -\frac{2\hat\mu^2}{\hat{m}})  \
e^{0} = (1 -\frac{4\mu^2}{m_\pi^2})e^{0}, 
\ee
where we made use of the Gell-Mann$-$Oakes$-$Renner relation. 
This is precisely what we find from
the general argument given in  previous section. To see this we evaluate the
two partition functions using (\ref{gen1}). First, note that 
the particle content of the phase quenched partition function is
a charged Goldstone boson  
and its anti-particle and two neutral Goldstone bosons. For $\mu < m_\pi/2$, 
their masses are given by $  m_\pi -2\mu$, $m_\pi + 2\mu$, $m_\pi$, $m_\pi$,
respectively, and the free energy is equal to by $2m \Sigma$. The Jacobian 
in the integration measure is a constant. The ordinary two flavor partition
function is of course obtained by setting $\mu=0$ in the expression 
obtained for the phase quenched partition function.   

\vspace{2mm}

For $\mu>m_\pi/2$ the thermodynamic limit of (\ref{pq-micro}) allows us to
determine both the free energy and the exponential prefactor. In this case   
the saddle point is inside the integration region and the
saddle point approximation to the partition function (\ref{pq-micro}) is given by
\be
Z_{1+1^*}(\hat{m};\hat\mu) \sim 
\sqrt{\frac 1{2\pi}} \frac 1 {\hat{\mu} \hat{m}}
e^{\hat m^2/2\hat\mu^2+2\hat\mu^2}
\qquad {\rm for} \qquad \hat m < 2\hat\mu^2.
\label{pq-lowm}
\ee
Notice that the prefactor of the exponential is larger by a factor of 
$\sqrt V$ than for $\mu<m_\pi/2$. The reason is that the Goldstone mode with
mass $m_\pi - 2\mu$ becomes exactly massless for $2\mu>m_\pi$. 
To use the saddle point approximation one must introduce a pion condensate
source term which lifts this mass away from zero 
(see Appendix \ref{app:jmass}).

The two flavor partition function remains $\mu$ independent so that in the
thermodynamic limit the average phase factor is given by
\be
\langle e^{2i\theta}\rangle_{1+1^*} \sim \left\{\begin{array}{ccc}
    (1 -\frac{4\mu^2}{m_\pi^2})  \ e^{0} & {\rm
      for} & m_\pi>2\mu, \\  & & \\
 \frac {1}{\sqrt{2 \pi V F_\pi^4}}\frac{2 \mu F_\pi}{m_\pi^2}
e^{-2VF_\pi^2\mu^2(1-m^2_\pi/4\mu^2)^2}
& {\rm      for} & m_\pi<2\mu .\end{array}\right.
\label{exp-termo-n=1}
\ee

Below we will derive the exact microscopic expression for the average phase
factor in the quenched and unquenched theories. To perform the calculations 
we will make use of recently developed random matrix techniques.

\vspace{2mm}

\section{The Random Matrix Model}
\label{sec:rmt}

In order to compute the average phase factor in the microscopic limit we use a
random matrix approach. In this limit the mass and chemical potential 
dependence of the QCD partition function is given by a random matrix 
partition function (see \cite{VW} for a review).
The expectation value of the phase factor is then given by the
ratio of two partition functions 
\be
\langle e^{2i\theta} \rangle_{N_f} = \frac{{Z}_{N_f+1|1^*}(m;\mu)}
{{Z}_{N_f}(m;\mu)},
\label{exp2ithZoverZ}
\ee
where, in general,  the random matrix partition function with $N_f$ quark flavors of 
mass $m_f$ and one pair of a regular quark and a conjugate bosonic quark both
with mass $m$ is defined by 
\be
{Z}_{N_f+1|1^*}(\{m_f\},m ;\mu) &\equiv&
 \int d\Phi  d\Psi \ w_G(\Phi)  w_G(\Psi) 
\prod_{f=1}^{N_f}{\det}({\cal D}(\mu) + m_f) \ 
\frac {{\det}({\cal D}(\mu) + m)}{{\det}({\cal D}^\dagger(\mu) + m)} .
\label{ZNfNb}
\ee
In the computation of the average phase factor we always take the quark masses 
to be degenerate. The Dirac operator is given by
\be
\label{dnew}
\mathcal{D}(\mu) = \left( \begin{array}{cc}
0 & i \Phi + \mu \Psi \\
i \Phi^{\dagger} + \mu \Psi^{\dagger} & 0
\end{array} \right) ~.
\ee
Here, $\Phi$ and $\Psi$ are complex $N\times N$  matrices with the same
Gaussian weight function 
\be
\label{wg}
w_G(X) ~=~ \exp( - N  \tr  X^{\dagger} X  ) ~.
\ee 

The partition function ${Z}_{N_f}(m;\mu)$ with $N_f$ flavors of 
mass $m$ is given by (\ref{ZNfNb}) without the ratio of the two 
determinants multiplying the $N_f$ flavor quark determinant. 
The microscopic limit
of the random matrix model is obtained by taking the size of the matrices
infinitely large, $N\to\infty$, while keeping the products 
\be
\hat{m}=2mN \ \ \ {\rm and} \ \ \ \hat{\mu}^2=2\mu^2N 
\ee 
fixed. Expressions involving the microscopic partition function 
are denoted by explicitly writing the
microscopic variables $\hat \mu$ and $\hat m$.  
\vspace{2mm}

The random matrix partition function
(\ref{ZNfNb}) can be rewritten in terms of an integral over the eigenvalues
of ${\cal D}$ and the unitary matrices that diagonalize ${\cal D}$. The
integral over the unitary matrices can be performed analytically \cite{O}
resulting in the eigenvalue representation of the partition function
\be
\label{epfnew}
{ Z}_{N_f+1|1^*}(\{m_f\},m;\mu)  \sim
 \int_{\mathbb{C}} \prod_{k=1}^{N} d^2z_k \,
{\cal P}_{N_f+1|1^*}(\{z_i\},\{z_i^*\}; \mu),
\ee
where the  integration extends over the full complex plane. 
The joint probability distribution of the eigenvalues is given by
\be
\label{jpd}
{\cal P}_{N_f+1|1^*}(\{z_i\},\{z_i^*\};\mu)
&=& \frac{1}{\mu^{2N}}\left|\Delta_N(\{z_l^2\})\right|^2 \, 
\prod_{k=1}^{N} w(z_k,z_k^*;\mu) \prod_{f=1}^{N_f}(m^2_f-z^2_k ) 
\frac{(m^2-z_k^2 )}
{(m^{2}-z_k^{*\,2}) }.
\ee 
The Vandermonde determinant is defined as  
\be
\Delta_N(\{z^2_l\}) \equiv \prod_{i>j=1}^N (z_i^2-z_j^2),
\label{vander}
\ee
and the weight function reads \cite{O}
\be 
w(z_k,z^*_k;\mu) &=& |z_k|^{2} 
K_0 \left( \frac{N (1+\mu^2)}{2 \mu^2} |z_k|^2 \right)
\exp\left(-\frac{N (1-\mu^2)}{4 \mu^2}  
(z^2_k + \conj{z_k}^2) \right). 
\label{wnew}
\ee

The eigenvalue representation makes it possible to define  
orthogonal polynomials in the complex plane 
\cite{fyodorov,akemann,AFV,BI,BII,O,AP}.
In order to evaluate the average phase factor we will make
use of such orthogonal polynomials and their Cauchy transform. 
The complex Laguerre polynomials given by \cite{O}
\be
p_k(z;\mu) = \left( \frac{1-\mu^2}N\right )^k k! \,
L_k \left ( -\frac{Nz^2}{1-\mu^2} \right)
\ee 
are the orthogonal polynomials corresponding to the weight $w(z,z^*;\mu)$
given in (\ref{wnew}). 
They satisfy the orthogonality relation
\be
\int_{\mathbb C}d^2z\ w(z,z^*;\mu)\ p_k(z;\mu)\ p_l(z;\mu)^* ~ 
 \ =\ r_k ~ \delta_{kl} \ ,
\label{OPdef}
\ee
with the norm 
\be
\label{Norm}
r_k ~=~
\frac{  \pi \, \mu^2 ~ (1+\mu^2)^{2k} ~ k! ~ k!}
     { N^{2k +  2}}  ~.
\ee
The Cauchy transform of the orthogonal polynomials is defined as
\be
h_k(m;\mu) = \int_{\mathbb C} d^2z \frac 1{z^2-m^2}w(z,z^*;\mu) p^*_k(z;\mu),
\ee
where ${\mathbb C}$ indicates that the integration extends over the complex
plane.

\section {The average phase factor in the microscopic limit}
\label{sec:exp2ith_micro}

Partition functions that are given by expectation values of ratios
of determinants can be expressed simply in terms of orthogonal 
polynomials and their Cauchy transforms 
\cite{fyodorov,akemann,AFV,BI,BII,AP,AOSV}. Generalizing
the results of \cite{AP} we find that the partition function
(\ref{ZNfNb}) for 
matrix size $N$ is given by
\be
{\cal Z}_{N_f+1|1^*}(\{m_f\},m;\mu) =
\frac{1}{r_{N-1}\Delta_{N_f+1}(\{m_f^2\})}  
 \left|\begin{array}{lll}
h_{N-1}(m;\mu) & \cdots & h_{N+N_f}(m;\mu)\\ 
p_{N-1}(m) & \cdots & p_{N+N_f}(m)\\
p_{N-1}(m_1) & \cdots & p_{N+N_f}(m_1)\\
\vdots & & \vdots \\
p_{N-1}(m_{N_f}) & \cdots & p_{N+N_f}(m_{N_f})\\
\end{array}
\right|.
\label{hpZNf}
\ee   
In order to make contact with QCD we take the microscopic limit of this
result. 
The orthogonal polynomials corresponding to the weight (\ref{wnew}) 
are the complex Laguerre polynomials.
In the microscopic limit, $\hat z = 2Nz$ for $N\to \infty$, the
polynomial $p_{N-1}/r_{N-1}$ is
\be 
\frac{p_{N-1}(\hat{z};\hatmu)}{r_{N-1}} 
&\sim & 
e^{-2 \hat\mu^2} I_0(\hat{z}),
\ee
where we have adopted the normalization used in \cite{SVbos}. 
Taking the limit of all quark masses equal to $m$ we obtain in the
microscopic limit (also in the remainder of the paper
the quark masses will be taken equal to $m$, which is indicated
by the notation)
\be
{Z}_{N_f+1|1^*}(\hat{m};\hat\mu) =
\frac 1{2^{N_f+1}} \frac{1}{\hat{m}^{N_f(N_f+1)}} 
\left|\begin{array}{lll}
X^{(0)}(\hat{m};\hat\mu) & \cdots & 
X^{(N_f+1)}(\hat{m};\hat\mu)\\ 
I_0(\hat{m}) & \cdots & \delta_{\hat{m}}^{N_f+1} I_0(\hat{m})\\ 
\vdots & & \vdots \\
\delta_{\hat{m}}^{N_f} I_0(\hat{m}) & \cdots & \delta_{\hat{m}}^{2N_f+1} I_0(\hat{m})\\
\end{array}
\right|.
\label{hpZNfmicro}
\ee   
Here, $\delta_{\hat{m}}\equiv\hat{m}\frac d {d\hat m}$ and 
\be
X^{(k)}(\hat{m};\hat\mu) \equiv -\frac 1{4\pi} 
\ \frac{1}{\hat{\mu}^2}e^{-2\hat{\mu}^2} 
\int_{\mathbb C} {\rm d}^2z \  
\frac{w(z,z^*,\hat{\mu}) \delta_{z^*}^k I_0(z^*)}{z^2-\hat{m}^2}. 
\label{xkdf}
\ee
This result was first presented  in the letter \cite{exp2ith-letter}. On the
next pages we examine this result in detail.

Denoting the  integrand of the $X^{(k)}$ by $F^{(k)}(z,z^*)$, we conclude
from the properties of Bessel functions that 
$[F^{(k)}(z,z^*)]^*=F^{(k)}(z^*,z)$. Therefore, the imaginary part
of the $X^{(k)}$ vanishes after integration over $d^2z$.

By using the identity ($z=x+iy$)
\be
I_k(z^*) = \frac {(-1)^{k}}{\pi}\frac{\sqrt{ z^*}}{\sqrt{-z^*}} 
(K_k(z^*) - (-1)^k K_k(-z^*)),\qquad k =0, 1, 2, \cdots\, ,
\label{bessident}
\ee
we can perform the $y$-integral in (\ref{xkdf}) by a contour integral
in the complex $y$-plane. By changing variables $z \to -z$ and
$z^* \to - z^*$  it can be easily shown that both terms in (\ref{bessident})
give the same integral. Our result is thus given by
\be
X^{(k)}(\hat{m};\hat\mu) = 
 \frac 1{2\pi^2} \ \frac{1}{\hat{\mu}^2}e^{-2\hat{\mu}^2}
\int_{-\infty}^\infty dx \int_{-\infty}^\infty dy 
\frac {|z|^{2}}{z^2-\hat m^2 }
K_0 \left( \frac{|z|^2}{4 \hat \mu^2}  \right)
\exp\left(-\frac{ (z^2 + \conj{z}^2)}{8\hat\mu^2} \right)
\delta_{z^*}^k   [ \frac{\sqrt{ z^*}}{\sqrt{-z^*}}  K_0(-z^*)]. 
\label{xk}
\ee
The only nonanalyticities in the complex $y$-plane are on the imaginary
$y$-axis, and for $|y| \to \infty$,  the integrand vanishes sufficiently fast 
in the lower part of the complex $y$-plane to deform the integration contour
as
\be
\int_{-\infty}^{-\epsilon} dy \cdots &\to& 
-\int_{-\epsilon}^{-\epsilon-i\infty} dy \cdots\,  ,
\nn\\
\int_{\epsilon}^{\infty} dy \cdots &\to& 
\int_{\epsilon}^{\epsilon-i\infty} dy \cdots \,.
\label{contours}
\ee
We integrate over the new contour by parameterizing $y =-is\pm \epsilon$
 which gives
an additional $-i$ from the Jacobian. The pole term can be decomposed into
a principal value part and a $\delta$-function,
\be
\frac 1{z^2-\hat{m}^2} 
&=& \frac 1{2\hat{m}}\left( \frac 1{z-\hat{m}} - \frac 1{z+\hat{m}} \right )
\nn\\ &=&
\frac 1{2\hat{m}} \left [\frac{x+s-\hat{m} \mp i\epsilon}{(x+s-\hat{m})^2 + \epsilon^2}
-\frac{ x+s+\hat{m}  \mp i\epsilon }{(x+s+\hat{m})^2 + \epsilon^2} \right ].
\label{pvdecom}
\ee
The principal value part combines with the discontinuity accross the
negative imaginary $y$-axis of the
other factors in the integrand in (\ref{xk}),
\be
X^{(k)}_{\rm cut}(\hat m; \hat\mu) &=& 
 \frac {i}{2\pi^2} \ \frac{1}{\hat{\mu}^2}
e^{-\frac{\hat m^2}{8\hat \mu^2}}e^{-2\hat{\mu}^2}
\int_{0}^\infty ds \int_{-\infty}^\infty dx  |z|^2
\frac 1{2\hat{m}}\left [    {\cal P} \frac 1{x+s+\hat{m}}
- {\cal P} \frac 1{x+s-\hat{m}}\right ]
\nn \\ && \times
{\rm Disc }\left[ K_0 \left( \frac{|z|^2}{4 \hat \mu^2}  \right)
\exp\left(-\frac{ (z^2 + \conj{z}^2)}{8\hat\mu^2} \right)
\delta_{z^*}^k   [ \frac{\sqrt{ z^*}}{\sqrt{-z^*}}  K_0(-z^*)] \right ]. 
\label{xkdisc}
\ee
Using the orthogonality of the polynomials on which this integral is based,
it can be shown that this integral vanishes \cite{OSV-new} 
\be
X^{(k)}_{\rm cut}(\hat m; \hat\mu) = 0.
\ee
What remains
is the $\delta$-function  part of (\ref{pvdecom}) in combination with the part
of the of
factors in the integrand that is continuous accross the negative imaginary
$y$-axis. By inspection one easily finds that the contribution of
the pole at $s = -x-\hat{m}$ vanishes. For the contribution of the pole
at $s= x-\hat{m}$ we have to distinguish $\hat{m}-x<x$ and $\hat{m}-x>x$. 
Because $\sqrt{z^*}/\sqrt{-z^*}$ has a cut for all values of $s$ (notice
that $z^* = x-s \mp \epsilon$), in the first case
the nonvanishing contribution comes from the cut in $K_0(|z|^2)$  and
in the second case from the cut in $K_0(-z^*)$. We thus find 
\be
X^{(k)}(\hat m; \hat\mu) &=& 
-\frac 1{4\hat\mu^2}
e^{-\frac{\hat m^2}{8\hat \mu^2}}e^{-2\hat{\mu}^2}
\left[ 
\int_{-\infty}^{\hat{m}/2} dx (2x-\hat{m})
I_0 \left( \frac{ (2x-\hat{m})\hat{m}}{4 \hat \mu^2}  \right)
\exp\left(-\frac{(\hat{m}-2x)^2}{8\hat\mu^2} \right)
\delta_{\hat{m}-2x}^k   [  K_0(\hat{m}-2x)]\right . \nn \\  
&& \left . -\int_{\hat{m}/2}^{\hat{m}} dx (2x-\hat{m})
K_0 \left(\frac{ (2x-\hat{m})\hat{m}}{4 \hat \mu^2}  \right)
\exp\left(-\frac{(2x-\hat{m})^2}{8\hat\mu^2} \right)
\delta_{2x-\hat{m}}^k   [  I_0(2x-\hat{m})] \right ]. 
\label{xkdisc2}
\ee
This result can be simplified to
\be
X^{(k)}(\hat m; \hat \mu )&=&
e^{-2\hat{\mu}^2}\frac{1}{4\hat{\mu}^2} e^{-\frac{\hat{m}^2}{8\hat\mu^2}}\nn
\\ && \times \left [
\int_{0}^{\hat{m}} du u \exp[ -\frac {u^2}{8\hat\mu^2}]
K_0\left ( \frac{u\hat{m}}{4\hat\mu^2}\right )(u\del_u)^k I_0(u) 
+\int_{0}^{\infty} du u \exp[ -\frac {u^2}{8\hat\mu^2}]
I_0\left 
( \frac{u\hat{m}}{4\hat\mu^2}\right )(u\del_u)^k K_0(u) \right ]. \label{X0}
\ee
Making use of the identities derived in Appendix \ref{app:id} these integrals 
can be rewritten as the sum of a polynomial in $\mu^2$ and an integral with
an essential singularity at $\mu = 0$,
\be
X^{(0)}(\hat m; \hat \mu )&=& K_0(\hat{m})
-e^{-2\hat{\mu}^2}\frac{1}{4\hat{\mu}^2} 
e^{-\frac{\hat{m}^2}{8\hat\mu^2}} 
\int_{\hat{m}}^\infty du u\exp[ -\frac {u^2}{8\hat\mu^2}]
K_0\left ( \frac{u\hat{m}}{4\hat\mu^2}\right ) I_0(u), 
\label{X0-final}
\\
X^{(1)}(\hat m; \hat \mu )&=& -\hat{m} K_1(\hat{m}) + 
4 \hat\mu^2 K_0(\hat{m})-
 e^{-2\hat{\mu}^2}\frac{1}{4\hat{\mu}^2} e^{-\frac{\hat{m}^2}{8\hat\mu^2}} 
\int_{\hat{m}}^\infty du u\exp[ -\frac {u^2}{8\hat\mu^2}]
K_0\left ( \frac{u\hat{m}}{4\hat\mu^2}\right )u\del_u I_0(u), 
\label{X1-final}
\\
X^{(2)}(\hat{m};\hat\mu) &=& K_0(\hat{m}) (\hat{m}^2 + 16 \hat\mu^4 
+ 8 \hat\mu^2) - 
  8 K_1(\hat{m})\hat{m} \hat\mu^2  \nn\\  &&
- e^{-2\hat{\mu}^2}\frac{1}{4\hat{\mu}^2} 
e^{-\frac{\hat{m}^2}{8\hat\mu^2}} 
\int_{\hat{m}}^\infty du u\exp[ -\frac {u^2}{8\hat\mu^2}]
K_0\left ( \frac{u\hat{m}}{4\hat\mu^2}\right )(u\del_u)^2 I_0(u), 
\label{X2-final}\\
X^{(3)}(\hat{m};\hat\mu) &=& 
K_0(\hat{m})(12\hat{m}^2\hat\mu^2+64\hat\mu^6+96\hat\mu^4
+ 2 \hat{m}^2 + 16 \hat\mu^2)
- K_1(\hat{m})(48\hat{m}\hat\mu^4 + \hat{m}^3 + 24\hat{m}\hat\mu^2)\nn\\  && \
- e^{-2\hat{\mu}^2}\frac{1}{4\hat{\mu}^2} 
e^{-\frac{\hat{m}^2}{8\hat\mu^2}}
\int_{\hat{m}}^\infty du u\exp[ -\frac {u^2}{8\hat\mu^2}]
K_0\left ( \frac{u\hat{m}}{4\hat\mu^2}\right )(u\del_u)^3 I_0(u), 
\label{X3-final}
\ee
The analytic properties of the $X^{(k)}$ will be discussed further in section 
\ref{sec:lattice}.

In oder to complete the computation of the average phase factor 
(\ref{exp2ithZoverZ}) we only need to recall that the ordinary 
($\mu$ independent) microscopic partition function is given by \cite{LS}
\be
Z_{N_f}(\hat{m})\sim \hat m^{-N_f(N_f-1)}\det[\delta_{\hat
  m}^{k+l}I_0(\hat{m})]_{k,l=0,\ldots,N_f-1}. 
\label{zmu0}
\ee
In the next section we will discuss explicitly the result for 
the average phase factor for $N_f=0,1$ and $2$.

\section{Explicit results and the thermodynamic limit of the phase factor}
\label{sec:exp2ith_ther}

In this section we will take a closer look at several specific cases and 
derive the large $\hat{m}$ and large $\hat\mu$ asymptotic expansions of 
the average phase factor from the exact microscopic expressions given
in the previous section. We refer to this as the thermodynamic limit 
of the microscopic results. These results confirm the expressions
for the average phase factor obtained in section \ref{sec:gen} from the mean
field argument.  
For the technical details of the asymptotic expansion
we refer to Appendix \ref{app:asymp}.  

\subsubsection{The Quenched Case}

We first consider the expectation value of the phase factor in the quenched
case. Then
\be
 \langle e^{2i\theta} \rangle_{N_f =0} & = &  \left | \begin{array}{cc}
X^{(0)}(\hat m; \hat \mu) & X^{(1)}(\hat m; \hat \mu)\\
I_0(\hat m) & \hat{m} I_1(\hat m) \end{array} \right | 
\\
&=& 1-4\hat\mu^2I_0(\hat{m})K_0(\hat{m}) \nn\\
&&-e^{-2\hat{\mu}^2}\frac{1}{4\hat{\mu}^2} e^{-\frac{\hat{m}^2}{8\hat\mu^2}} 
\int_{\hat{m}}^\infty dx x \exp[ -\frac {x^2}{8\hat\mu^2}]
K_0\left ( \frac{x\hat{m}}{4\hat\mu^2}\right ) \left(I_0(x)\hat{m}
I_1(\hat{m})-x I_1(x)I_0(\hat{m})\right),
\label{quenched-micro}
\ee
where we made use of the Wronskian identity
$\hat{m}I_0(\hat{m})K_1(\hat{m})+\hat{m}I_1(\hat{m})K_0(\hat{m})=1$.  

We now take the thermodynamic limit $\hat{m}\to\infty$ and
$\hat\mu\to\infty$. 
In the normal phase we can use the leading order asymptotic expansions
for the $X^{(k)}$ and the Bessel functions whereas in the condensed phase
we have to include the subleading corrections. Substituting the expressions
obtained in Appendix \ref{app:asymp} we find (the result for $m_\pi <2\mu$
is not valid very close to the critical point (see Appendix C))
\be
\langle e^{2i\theta}\rangle_{N_f=0} \sim \left\{\begin{array}{ccc}
    (1-\frac{4\mu^2}{m_\pi^2})\ e^{0} & {\rm
      for} & m_\pi>2\mu, \\ &&\\
 \frac 1{V^{3/2}\sqrt{2\pi}}\frac{1}{m_\pi^2 F_\pi^3\mu}\frac 1{2(1-m_\pi^2/4\mu^2)^2} \ 
e^{-2VF_\pi^2\mu^2(1-m^2_\pi/4\mu^2)^2} & {\rm
      for} & m_\pi<2\mu .\end{array}\right.
\label{exp-quen-thermo}
\ee   
For $m_\pi < 2\mu$ the leading order terms contributing to the
prefactor cancel. This asymptotic result agrees with the result 
we have obtained  from the general arguments in section \ref{sec:gen}.

\subsubsection{Full QCD for $N_f =1$}

For one flavor, the explicit microscopic expression for the expectation value of
the phase factor is given by
\be
\langle e^{2i\theta}\rangle_{N_f=1} \ 
& = & \frac{1}{2\hat{m}^2 I_0(\hat{m})}
 \left|\begin{array}{lll}
X^{(0)}(\hat{m}^*;\hat\mu) & X^{(1)}(\hat{m}^*;\hat\mu) & X^{(2)}(\hat{m}^*;\hat\mu)\\ 
I_0(\hat{m}) & \hat{m} I_1(\hat{m})& \hat{m}^2 I_0(\hat{m})\\ 
\hat{m} I_1(\hat{m}) &  \hat{m}^2 I_0(\hat{m}) &
\hat{m}^2(2 I_0(\hat{m})+\hat{m}I_1(\hat{m})) \\
\end{array}
\right|,
\label{exp-Nf1-micro}
\ee
where the $X^{(k)}$ have been given in the previous section and we used that
$Z_{N_f=1} (\hat m;\hat\mu) = I_0(\hat m)$. 
With the expressions found for $X^{(k)}$ the 
average phase factor can be written as 
\be 
\langle e^{2i\theta}\rangle_{N_f=1} & = & 1
-4\left(\hat{m}
K_0(\hat{m})I_0(\hat{m})+\frac{I_1(\hat{m})^2}
{I_0(\hat{m})}(K_0(\hat{m})-\hat{m}K_1(\hat{m}))\right)\hat\mu^2+8K_0(\hat{m})
\frac{I_0(\hat{m})^2-I_1(\hat{m})^2}{I_0(\hat{m})}\hat\mu^4\nn\\
&&\hspace{-1cm}
-\frac{e^{-2\hat{\mu}^2-\frac{\hat{m}^2}
{8\hat\mu^2}}}{8\hat{m}^2\hat{\mu}^2 I_0(\hat{m})}
\int_{\hat{m}}^\infty dx x \exp[ -\frac {x^2}{8\hat\mu^2}]
K_0\left ( \frac{x\hat{m}}{4\hat\mu^2}\right )
 \left|\begin{array}{lll}
I_0(x) & \delta_x I_1(x)& (\delta_x)^2 I_0(x)\\
I_0(\hat{m}) & \delta_{\hat{m}} I_0(\hat{m})& 
(\delta_{\hat{m}})^2 I_0(\hat{m})\\
\delta_{\hat{m}} I_0(\hat{m}) &  (\delta_{\hat{m}})^2 I_0(\hat{m}) &
(\delta_{\hat{m}})^3  I_0(\hat{m}) 
\\ \end{array} \right|. \label{exp-Nf1-micro-v2} \ee  

The large $\hat m$ and $\hat \mu$ limit of the expectation value of the 
phase follows from the asymptotic expressions for the $X^{(k)}$ given
in Appendix \ref{app:asymp} and the asymptotic expressions for the Bessel functions
\be
\langle e^{2i\theta}\rangle_{N_f=1} \sim  \left\{\begin{array}{ccc}
    (1 -\frac{4\mu^2}{m_\pi^2})^2  \ e^{0} & {\rm
      for} & m_\pi>2\mu \\  & & \\
O(\frac 1{V^{5/2}})
e^{-2VF_\pi^2\mu^2(1-m^2_\pi/4\mu^2)^2}
& {\rm
      for} & m_\pi<2\mu \ . \end{array}\right. 
\label{exp-termo-Nf1}
\ee  
The first three orders contributing to the 
 prefactor of the result for $m_\pi<2\mu$ 
vanish.

Since the average phase factor is the ratio of two partition functions
it is necessarily real even if the ``statistical weight'' is complex
such as for $N_f = 1$. This does not imply that 
$\langle \sin 2\theta \rangle$ vanishes such as in the quenched theory or
the phase quenched theory. Since $\langle  \sin 2\theta \rangle =
\langle \exp(2i\theta ) - \exp(-2i\theta) \rangle/2i$ and
\be
\langle e^{-2i\theta} \rangle_{N_f =1} = 
\frac{Z_{N_f=1^*}(m;\mu)}{Z_{N_f=1}(m;\mu)} =1,
\ee
we have in the microscopic limit that
\be
\langle \sin 2\theta \rangle_{N_f =1} = \frac 1{2i} 
\left [ \langle e^{2i\theta} \rangle_{N_f=1} -1 \right ].
\ee
We thus find that the expectation value 
$\langle \sin 2\theta \rangle_{N_f =1}$ is generally nonzero
and  purely imaginary. Its thermodynamic limit is given by
\be
\langle \sin(2\theta) \rangle_{N_f=1} \sim  
\left\{\begin{array}{ccc} \frac{1}{2i}
\big((1-\frac{4\mu^2}{m_\pi^2})^2-1\big) & {\rm
      for} & m_\pi>2\mu, \\ && \\ -\frac{1}{2i} & {\rm
      for} & m_\pi<2\mu .\end{array}\right.
\label{sin-termo-Nf1}
\ee    

Since the average of the inverse phase factor is 1 we automatically also
know the variance of the phase factor
\be
 \langle e^{2i\theta} e^{-2i\theta} \rangle_{N_f=1}-\langle
 e^{2i\theta}\rangle_{N_f=1} \langle e^{-2i\theta} \rangle_{N_f=1}
=1-\langle e^{2i\theta}\rangle_{N_f=1}.
\ee 
So while the average phase factor goes to zero, its variance goes
to 1 for $\mu>m_\pi/2$, what would also have been the result
for a uniformly random distribution of the phase.
This central feature of the sign problem 
is also present for larger number of flavors as well as for 
quenched and phase quenched QCD (in the latter two cases there is 
of course no sign problem).

\subsubsection{Two dynamical flavors}

With two flavors the explicit microscopic expression for the average phase
factor is  
\be
\langle e^{2i\theta}\rangle_{N_f=2} &= &
\frac{1}{8\hat{m}^6 (I_0(\hat{m})^2-I_1(\hat{m})^2)} \\
&&\hspace{-1cm}\times  \left|\begin{array}{llll}
X^{(0)}(\hat{m}^*;\hat\mu) & X^{(1)}(\hat{m}^*;\hat\mu) &
X^{(2)}(\hat{m}^*;\hat\mu)& X^{(3)}(\hat{m}^*;\hat\mu)\\ 
I_0(\hat{m}) & \delta_{\hat m} I_0(\hat m) & 
(\delta_{\hat m})^2 I_0(\hat m) & (\delta_{\hat m})^3 I_0(\hat m)\\ 
\delta_{\hat m} I_0(\hat m) & (\delta_{\hat m})^2 I_0(\hat m) &
(\delta_{\hat m})^3 I_0(\hat m) & (\delta_{\hat m})^4 I_0(\hat m)\\ 
(\delta_{\hat m})^2 I_0(\hat m) & (\delta_{\hat m})^3 I_0(\hat m)& 
(\delta_{\hat m})^4 I_0(\hat m) & (\delta_{\hat m})^5 I_0(\hat m)
\end{array}
\right|. \nn
\label{exp-Nf2-micro}
\ee

\begin{figure}[!t]
  \unitlength1.0cm
    \epsfig{file=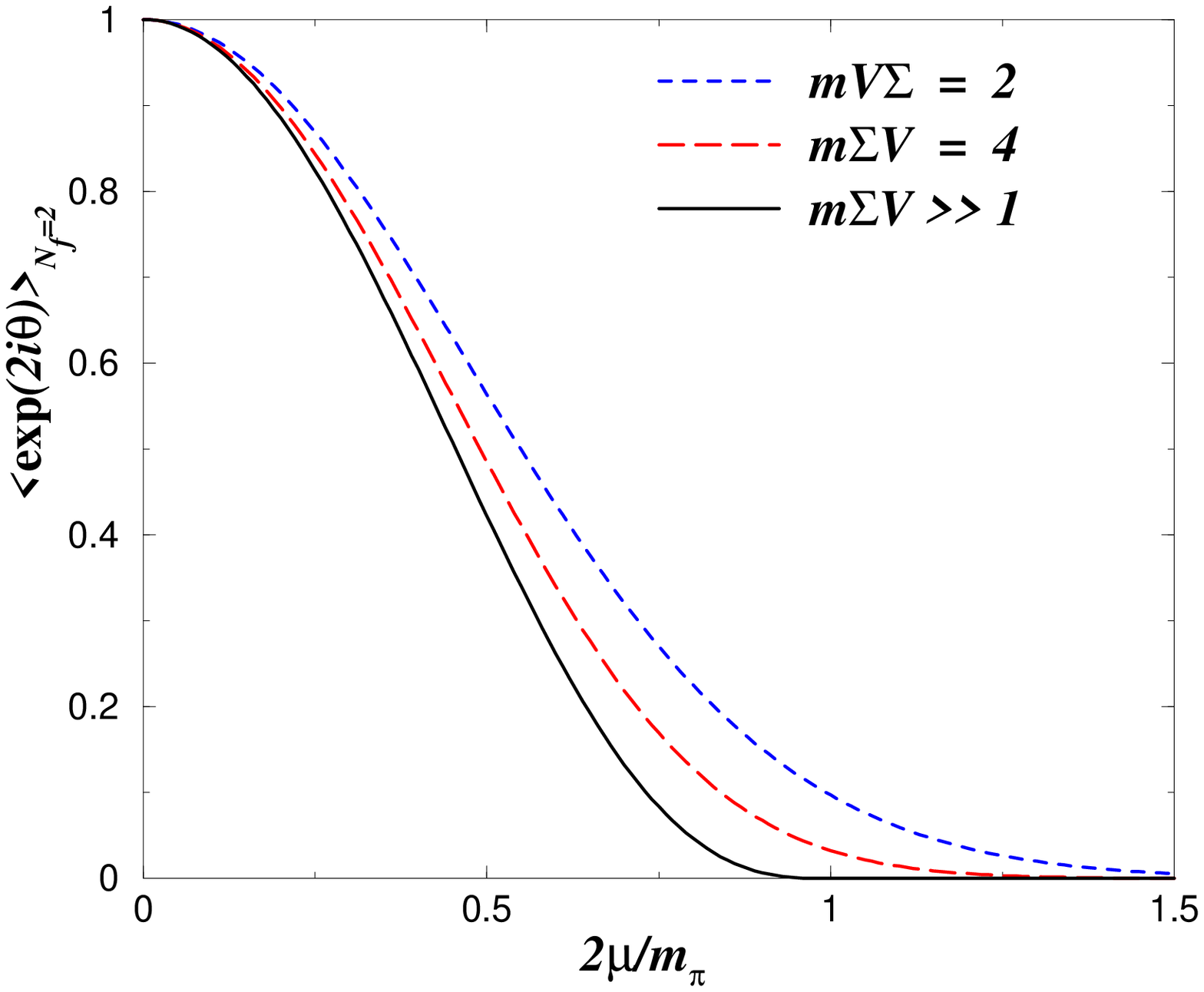,clip=,width=8cm}
    \epsfig{file=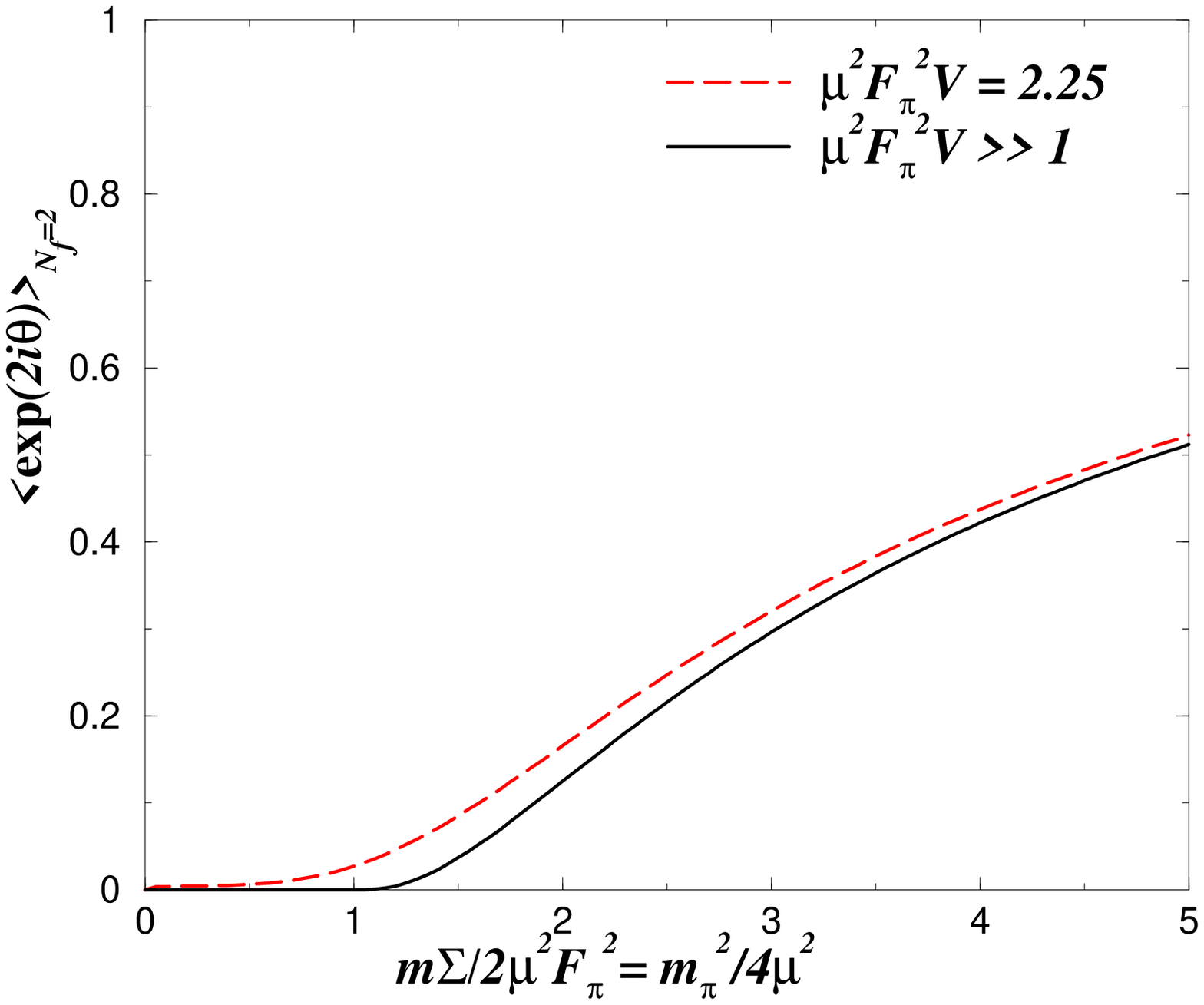,clip=,width=8cm}
  \caption{
  \label{fig:phase} {\bf Left:} The average phase factor for two dynamical
  flavors as a function of the chemical potential for fixed quark mass. The
  full curve represents the result in the thermodynamic limit. Note that the
  convergence to the 
  thermodynamic limit is particular rapid for small values of the chemical
  potential. {\bf Right:} The average phase factor as a function of the quark
  mass for fixed chemical potential. The full curve again displays 
the result in
  the thermodynamic limit.}
\end{figure}

For $\mu<m_\pi/2$ the thermodynamic limit of this result is  
\be
\langle e^{2i\theta}\rangle_{N_f=2} \sim 
    (1-\frac{4\mu^2}{m_\pi^2})^3 \ e^{0} \ \ \  {\rm
      for} \ \ \  m_\pi>2\mu.  
\label{exp-termo-Nf2-num}
\ee 
As the chemical potential increases beyond half the pion mass the average
phase factor is again exponentially small in the volume in agreement with
(\ref{exp2ith_therm_cond}). The fast convergence to the thermodynamic limit
is illustrated in figure \ref{fig:phase}.

\section{Analytic continuation from imaginary chemical potential}
\label{sec:lattice}

For purely imaginary chemical potential the fermion determinant is real, and
the phase factor is equal to unity. 
Thus it seems to be impossible  to calculate the 
average phase factor at imaginary values of $\mu$ and then perform an analytic
continuation in $\mu$. However, if we first use that
$\det(D+\mu\gamma_0+m)^*=\det(D-\mu\gamma_0+m)$ to write the average 
phase factor at 
real chemical potential as
\be
\left \langle e^{2i\theta(\mu)} \right \rangle_{N_f} = \left \langle
\frac{\det(D+\mu\gamma_0+m)}{\det(D-\mu\gamma_0+m)} \right \rangle_{N_f},
\label{exp2ithmu}
\ee
and subsequently substitute $\mu$ by $i\mu$ we obtain a real number
\be
\left \langle e^{2i\theta(i\mu)}\right \rangle_{N_f}  \equiv \left \langle
\frac{\det(D+i\mu\gamma_0+m)}{\det(D-i\mu\gamma_0+m)}\right \rangle_{N_f}
\label{exp2ithimu}
\ee
that is (typically) different from one. Below we will use this analytical
continuation between real and imaginary chemical potential. 
 
The microscopic limit of the 
quenched average of (\ref{exp2ithimu}) for imaginary 
chemical potential can be derived by means of the supersymmetric
method. It is given by \footnote{This result was
  obtained in collaboration with P.H. Damgaard. 
To derive it one adds $-\mu^2F_\pi^2V/4
{\rm Str}[\sigma_3,U][\sigma_3,U^{-1}]$ to the action in section 4 of
\cite{DOTV} and computes the partition function for equal masses.}
\be
\langle e^{2i\theta(i\mu)} \rangle_{N_f=0} 
= 1 + 4\hat\mu^2 I_0(\hat{m})K_0(\hat{m}).  
\label{avphaseimu}
\ee
Since the result is a polynomial in $\hat\mu^2$,  analytical continuation 
to real chemical potential is trivial: we simply flip the sign of $\mu^2$. 
This result should be contrasted with the result for real $\mu$ given in
Eq. (\ref{quenched-micro}):
\be
\langle e^{2i\theta(\mu)} \rangle_{N_f=0} 
&=& 1-4\hat\mu^2I_0(\hat{m})K_0(\hat{m}) 
-\frac{e^{-2\hat{\mu}^2}}{4\hat{\mu}^2} e^{-\frac{\hat{m}^2}{8\hat\mu^2}} 
\int_{\hat{m}}^\infty dx x e^{ -\frac {x^2}{8\hat\mu^2}}
K_0\left ( \frac{x\hat{m}}{4\hat\mu^2}\right ) \left(I_0(x)\hat{m}
I_1(\hat{m})-x I_1(x)I_0(\hat{m})\right)\nn \\
&\sim& 1 - 4\hat\mu^2 I_0(\hat{m})K_0(\hat{m}) 
+\hat\mu^2 K_0\left (\frac{\hat{m}^2}{4\hat\mu^2}\right ) 
e^{-\frac{\hat{m}^2}{4\hat\mu^2}-2\hat\mu^2}(I_0(\hat{m})^2-I_1(\hat{m})^2)
\quad {\rm for} \quad \hat\mu\to 0 .  
\label{quenched-microb}
\ee
What is reproduced by the expression (\ref{avphaseimu}) 
 for  purely imaginary chemical
potential are the analytic terms in $\mu$,  but the
term with the essential singularity is not obtained.
Note that the non-analytic term does not contribute to the Taylor series of 
(\ref{quenched-microb}). The Taylor expansion around $\mu=0$ of
(\ref{quenched-microb}) is thus given by the first two terms only.
This expansion can be analytically continued in $\mu$ and necessarily
reproduces the result for imaginary $\mu$ because its Taylor expansion
is finite.

The microscopic limit of the average phase factor for imaginary $\mu$ 
can also be evaluated for $N_f \ne 0$ 
using the supersymmetric method of \cite{DOTV}. 
However, it is possible to obtain 
$\langle\exp(2i\theta(i\mu))\rangle_{N_f}$ from the results 
at real $\mu$ without a detailed computation. 
We only need that the average phase factor for imaginary $\mu$ is given 
by a finite power series in $\mu^2$. This immediately follows from 
the super symmetric approach where $i\mu$ only occurs in combination with 
Grassmann variables. Therefore, 
also for $N_f \ne 0$, the exact average phase factor at imaginary $\mu$ can be
obtained from the analytical continuation of the Taylor expansion
of the expression for real $\mu$. For example, for $N_f =1$ we obtain the
result 
\be
\langle e^{2i\theta(i\mu)}\rangle_{N_f=1} 
= 
1+4\left(\hat{m}K_0(\hat{m})I_0(\hat{m})
+\frac{I_1(\hat{m})^2}{I_0(\hat{m})}
(K_0(\hat{m})-\hat{m}K_1(\hat{m}))\right)
\hat\mu^2+8K_0(\hat{m})
\frac{I_0(\hat{m})^2-I_1(\hat{m})^2}{I_0(\hat{m})}\hat\mu^4.
\ee

Conversely, if we would have calculated the ``phase factor'' for imaginary
chemical potential and then made the analytical continuation to real $\mu$ 
we only would have obtained the polynomial terms in $\mu$. Terms for which 
all derivatives at $\mu=0$ vanish cannot possibly be obtained from a 
Taylor expansion.

For $\mu \ll m_\pi/2$ the term with the essential singularity is not
important. However, it becomes the dominant term when $\mu$ approaches 
$m_\pi/2$. For $\mu > m_\pi/2$ the non-analytic term cancels the terms
that are polynomial in $\mu$, and is responsible for the exponential 
suppression of average phase factor. In the thermodynamic limit, the
non-analytic term has a phase transition at $\mu = m_\pi/2$. For $\mu < m_\pi/2$
the saddle point is outside the integration range and the leading contribution
to the integral comes from its lower limit. On easily shows that the 
integral vanishes for $V\to \infty$ and $\mu < m_\pi/2$. For $\mu >m_\pi/2$
the saddle point is inside the integration domain 
and its leading order exactly cancels the thermodynamic limit of the
terms that are polynomial in $\mu$.

A similar essential singularity at $\mu=0$ is found for $N_f =2$, and
the results obtained by Taylor expansion or
analytic continuation in $\mu$ only reproduce the finite polynomial series 
in $\hat\mu^2$. 
What is not reproduced by analytic continuation are 
 precisely the contributions given by the
integrals in (\ref{X0-final})-(\ref{X2-final}).

The integrals with the essential singularity are due to Dirac eigenvalues 
with real part larger than $m$. This is best illustrated by
considering $X^{(0)}$ which is the partition function for one bosonic
flavor \cite{SVbos}. In terms of a joint eigenvalue distribution it is given by
\be
Z_{N_f=-1}^N(m;\mu) 
&=& \frac{1}{\mu^{2N}} \int \prod_k d^2 z_k 
\left|\Delta_N(\{z_l^2\})\right|^2 \, 
\prod_{k=1}^{N} w(z_k,z_k^*;\mu)
\frac 1 {(m^2-z_k^2 )},
\ee 
where the Vandermonde determinant and the joint probability distribution 
are defined in (\ref{vander}) and (\ref{wnew}), respectively. 
If we single out one eigenvalue, which we will denote by $w$ 
this can be re-expressed as
\be
Z_{N_f=-1}^N(m;\mu) 
&=& \frac{N}{\mu^{2}}\int d^2 w
 |w|^{2} 
K_0 \left( \frac{N (1+\mu^2)}{2 \mu^2} |w|^2 \right)
\exp\left(-\frac{N (1-\mu^2)}{4 \mu^2}  
(w^2 + \conj{w}^2) \right)
\frac 1 {(m^2-w^2 )} Z_{N_f=1}^{N-1}(w^*),
\ee 
where we have used the identity \cite{AP}
\be
\prod_{k=1}^N \frac 1{m^2-z_k^2} = \sum_{j=1}^N \frac 1{m^2-z_j^2} 
\prod_{k\ne j}^N \frac 1{z_j^2 -z_k^2}.
\ee
Using that in the microscopic limit the partition 
function $Z_{N_f=1}^{N-1}(w^*) $
is given by
\be                           
Z_{N_f=1}^{N-1}(w^*) \to I_0(\hat w^*).
\ee
we obtain for the microscopic limit of $Z_{N_f=-1}^N$  (which as before 
is denoted by $X^{(0)}$ using the same normalization)
\be
\label{jpdzm}
X^{(0)}(\hat{m};\hat\mu)
&=& \frac 1{4\pi}\frac{1}{\hat \mu^{2}}\int d^2 \hat{w} 
 |\hat{w}|^{2} \frac 1 {(\hat m^2- \hat{w}^2 )}
K_0 \left( \frac{|\hat{w}|^2   }{2 \hat \mu^2} \right)
\exp\left(-\frac{ \hat{w} + \hat{w}^{*\,2}  }{4 \hat\mu^2}  
 \right)I_0( \hat{w}^*).
\ee 
The expression for $X^{(0)} $ given in (\ref{X0-final}) 
is obtained by integration 
over the imaginary part of $\hat{w}$ by means of Cauchy's theorem and 
introducing the variable $u= 2{\rm Re}(\hat{w}) - \hat{m}$.
The integral over $u>\hat{m}$ in (\ref{X0-final}), 
thus corresponds to contributions with 
the real part of  one
of the  eigenvalues larger than $m$. The same argument can be made
for $X^{(k>0)}(\hat m; \hat \mu)$.

\begin{figure}[!t]
  \unitlength1.0cm
    \epsfig{file=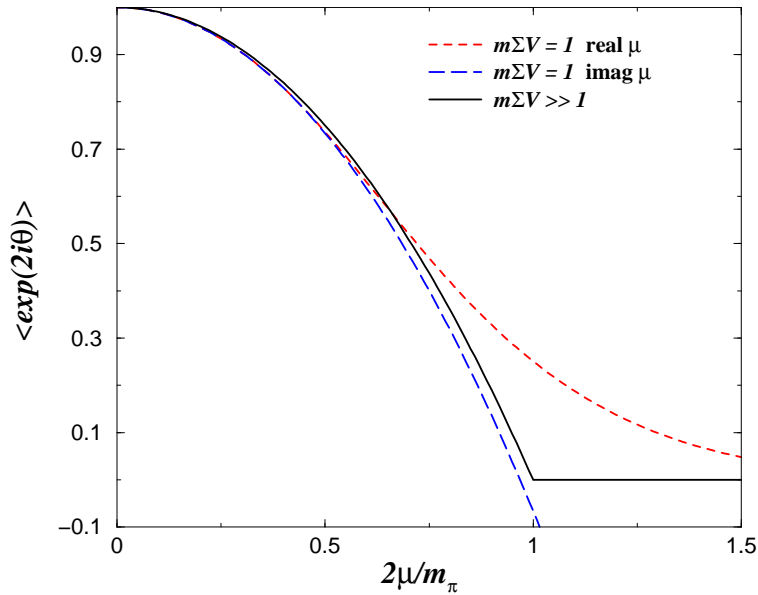,clip=,width=10cm}
  \caption{ \label{fig:imu} The quenched average of the phase factor. Shown
    is the microscopic result for real $\mu$ (upper short dashed curve) and
    the microscopic result obtained by analytic continuation from imaginary
    chemical potential (lower long dashed curve). The full curve represents
    the thermodynamic result for real $\mu$. Note that while the average
    phase factor is 
    non-analytic at $\mu=0$, the analytic continuation from imaginary $\mu$
    follows closely the correct result for $2\mu/m_\pi\ll 1$.}
\end{figure}

For $\hat\mu^2>\hat{m}/2$ the probability
that one of the eigenvalues has a real part larger than $m$ 
remains nonzero in the thermodynamic limit so that the non-analytic term
becomes important. In the thermodynamic limit it exactly cancels the 
analytic term resulting in a vanishing average phase factor.
In figure \ref{fig:imu} we illustrate that the contribution of the 
non-analytic term (given by the difference between the upper
and lower dashed curve) is important close to the critical value of $\mu$ and
beyond. 
For comparison we also give the thermodynamic limit of the average phase factor
(full curve).

The source of  the non-analyticity is the  inverse
determinant of a non-hermitian operator. For example, for the partition 
function with one bosonic flavor with
 imaginary chemical potential
the commutator term in the chiral Lagrangian vanishes resulting in
a microscopic partition function that is independent of $\mu$. 
For real $\mu$ the bosonic determinant has to be regularized resulting in 
a $\mu$-dependent effective Lagrangian (see \cite{SVbos} for details), and
a partition function that is $\mu$-dependent and therefore non-analytic at
$\mu=0$.

The message we would like to convey is that
the Taylor expansion in $\mu$ about zero of averages which involve an 
inverse quark determinant has zero radius of convergence. Also traces of the
inverse of non hermitian operators lead to essential singularities at $\mu=0$
unless the inverse is compensated by a factor from the fermionic determinant. 
The simplest example is
the quenched chiral condensate which is non-analytic in $\mu$ at $\mu=0$.

Note that the average phase factor in the phase quenched theory 
(\ref{exp2ithPQ}) does not involve averages of inverse determinants and is 
analytic in $\mu$ at $\mu=0$ as can easily be verified from (\ref{pq-phase}).

\section{1-loop corrections} 

So far we have focused on the average phase factor in the microscopic domain 
where $m_\pi^2F_\pi^2 \sim 1/V$ and $\mu^2F_\pi^2 \sim 1/V$ as the volume is
taken to infinity. A perturbative expansion scheme where $p \sim 1/L$,
$m_\pi F_\pi \sim  1/L^2$, $\mu F_\pi \sim 1/L^2$ is known as the 
$\epsilon$-expansion \cite{GLeps}. The microscopic domain is the leading
order term in the $\epsilon$-expansion. In this section we consider the so
called  $p$-expansion where $p \sim 1/L$, $m_\pi \sim  1/L$, $\mu
\sim 1/L$ and work  
out the average phase factor in an expansion in $m_\pi^2L^2$ 
and $\mu^2L^2$ for a finite but large box.
In the $p$-expansion, the 
microscopic variables $\mu^2F_\pi^2V$ and $m\Sigma V$ are large, and
consequently the leading term in the expansion is given by the asymptotic 
limit of the microscopic expression. Below we will compute 
subleading terms  
 up to order $m_\pi^2\mu^2L^4$ 
and $\mu^4L^4$. This corresponds to the 1-loop corrections generated 
by the nonzero momentum terms of the Goldstone bosons. For simplicity we 
will consider only the average phase factor in the phase quenched theory 
for $\mu<m_\pi/2$.

\subsection{1-loop Integral at $\mu \ne 0$}

In the standard notation of chiral perturbation theory we have 
\cite{HL}
\be
\langle e^{2i\theta}\rangle_{1+1^*}
=\frac{Z_{N_f=2}}{Z_{1+1^*}}=\exp(G_0(\mu=0)-G_0(\mu)).
\ee
Notice that the contribution of the neutral Goldstone bosons cancels, and
that we have a factor 2 in the exponent because the contribution
of the two oppositely charged Goldstone bosons is the same.
The 1-loop contributions of a single charged Goldstone boson with 
charge $2$ ($\mu$ is the quark chemical potential so a pion made from an up
and a down quark has charge 2) in a box with volume $L^d$ is given by
\be
e^{G_0(\mu)/2} \equiv \exp[-\frac 12 \sum_{p_{k\, \alpha}}
\log(\vec p^2_k +m^2_\pi +(p_{k\, 0}-2i \mu)^2) ],
\ee
where
\be
p_{k\,\alpha} = \frac {2\pi k_\alpha}{L_\alpha}, 
\qquad k_\alpha \quad {\rm integer}.
\ee
(For a fermionic Goldstone boson the sign of the exponent is positive.)
The contribution to the free energy given by $G_0(\mu)$ is divergent. 
However, the difference between $G_0(\mu)$ and $G_0(\mu)$ for $V=\infty$
is finite, that is 
\be
G_0(\mu) = G_0(\mu)|_{V=\infty} + g_0(\mu)
\ee
with $g_0(\mu)$ finite. Since $G_0(\mu)|_{V=\infty}$ is independent of
$\mu$ for $\mu <m_\pi/2$ (in dimensional regularization, see e.g.\cite{STV}) 
we have that  
\be
\langle e^{2i\theta}\rangle_{1+1^*}
=\frac{Z_{N_f=2}}{Z_{1+1^*}}=\exp(g_0(\mu=0)-g_0(\mu)).
\label{finite}
\ee
We wish to evaluate this 1-loop contribution
including the $1/V$ corrections. This can be 
done along the lines of \cite{HL} where the expression for $g_0$ was 
worked out for $\mu=0$. The first step is to perform a Poisson resummation
\be
G_0(\mu)= -\sum_{p_{ k\,\alpha }}
\log(\vec p^2_k +m^2_\pi +(p_{k\, 0}-2i \mu)^2) 
=-V \sum_{l_\alpha}\int \frac{d^d p}{(2\pi)^d } e^{iL_\alpha  p_\alpha l_\alpha} 
\log(\vec p^2 +m^2_\pi +(p_{0}-2i \mu)^2) ,
\ee
where the sum is over all integers. The thermodynamic limit,
$G_0|_{V=\infty}$, is given by the term $l_\alpha = 0$. 
Since this term does not contribute to $\langle\exp(2i\theta)\rangle$ 
we exclude it from now on. For $ \mu < m_\pi/2$ we can shift the $p_0$ 
variable by $2i\mu$. This results in
\be
g_0 &=&-
V {\sum_{l_\alpha}}'
\int \frac{d^d p}{(2\pi)^d } e^{iL_\alpha p_\alpha l_\alpha-2\mu l_0 L_0} 
\log(\vec p^2 +m^2_\pi +p_{0}^2)\nn \\
&=&-V {\sum_{l_\alpha}}'
\int \frac{d^d p}{(2\pi)^d } e^{iL_\alpha p_\alpha l_\alpha-2\mu l_0 L_0} 
 \lim_{\epsilon \to 0}\left[ -\gamma +\frac 1\epsilon - 
\int_0^\infty \frac{d\lambda}{\lambda^{1-\epsilon}}
e^{-(\vec p^2 +m^2_\pi + p_{0}^2)\lambda } \right ],
\ee
where $\gamma$ is the Euler constant.
The $1/\epsilon-\gamma$-terms do not contribute 
because the term with all $l_\alpha =0$
has been excluded from the sum. 
After performing the momentum integrals, 
the limit $\epsilon \to 0 $ can be taken safely resulting in
\be  
g_0
&=&
V {\sum_{l_\alpha}}' \frac{1}{(2\pi)^d } \pi^{d/2} 
e^{-2\mu l_0 L_0} 
\int_0^\infty \frac{d\lambda}{\lambda}{\lambda^{-d/2}}
e^{ -\frac{l_\alpha^2 L^2_\alpha}{4\lambda}} 
e^{-m^2_\pi\lambda }.
\ee
Next we change to dimensionless integration variables by
\be
\lambda \to \lambda \frac{L^2}{4 \pi}\qquad {\rm with} \qquad
L = (\prod_\alpha L_\alpha)^{1/d}
\ee
and interchange the sums and the integral to arrive at
\be
g_0
&=&
\int_0^\infty \frac{d\lambda}{\lambda}{\lambda^{-d/2}}
e^{-m^2_\pi L^2\lambda /4\pi }
(\prod_\alpha {\sum_{l_\alpha}} e^{-2\mu l_0 L_0\delta_{\alpha 0}} 
e^{ -\pi \frac{l_\alpha^2 L_\alpha^2}{\lambda L^2}} -1).
\ee
The integral over $\lambda$ can be split into into a  part with $\lambda < 1$
and a part with $\lambda >1$. On the second part we apply Jacobi's
imaginary transformation
\be
\sum_{k=-\infty}^\infty e^{-ak^2-bk} 
= \sqrt{\frac \pi a}
e^{\frac{b^2}{4a}}\sum_{k=-\infty}^\infty e^{-\frac{\pi^2 k^2}a-\frac{\pi i b k}a} .
\ee
This leads to 
\be
g_0
=
\int_0^1 \frac{d\lambda}{\lambda}{\lambda^{-d/2}}
e^{-m^2_\pi L^2\lambda /4\pi }
(\prod_\alpha {\sum_{l_\alpha}} e^{-2\mu l_0 L_0\delta_{\alpha 0}} 
e^{ -\pi \frac{l_\alpha^2 L_\alpha^2}{\lambda L^2}}-1) \nn\\
+   \int_1^\infty \frac{d\lambda}{\lambda}
e^{\frac{\mu^2 L^2\lambda } 
{\pi }}
e^{-m^2_\pi L^2\lambda /4\pi }
(\prod_\alpha {\sum_{l_\alpha}} 
e^{-2 i \mu l_0 \frac{\lambda L^2}{ L_0}\delta_{\alpha 0}} 
e^{ -\pi l_\alpha^2 \frac{\lambda L^2}{ L_\alpha^2}}-1)  +r_0.
\ee
The Jacobi imaginary transformation is applied to the full sum so that
the subtracted term before and after the transformation is different.
This difference, denoted by $r_0$, is given by
\be
r_0
= 
\int_1^\infty \frac{d\lambda}{\lambda}
e^{\frac{\mu^2 L^2  \lambda } {\pi}}
e^{-m^2_\pi L^2\lambda /4\pi } -
 \int_1^\infty \frac{d\lambda}{\lambda}{\lambda^{-d/2}}
e^{-m^2_\pi L^2\lambda /4\pi } .
\ee
Finally, we change $\lambda \to 1/\lambda$ in the Jacobi transformed
terms in $g_0$ to obtain
\be
g_0
=
\int_0^1 \frac{d\lambda}{\lambda}{\lambda^{-d/2}}
e^{-m^2_\pi L^2\lambda /4\pi }(\prod_\alpha
{\sum_{l_\alpha}} e^{-2\mu l_0 L_0\delta_{\alpha 0}} 
e^{ -\pi \frac{l_\alpha^2 L_\alpha^2}{\lambda L^2}}-1) \nn\\
+ \int_0^1 \frac{d\lambda}{\lambda}
 e^{\frac{\mu^2L^2 } 
{\pi\lambda }}
e^{-m^2_\pi L^2/(4\pi \lambda  )}(\prod_\alpha 
{\sum_{l_\alpha}} e^{-2 i \mu l_0 \frac{ L^2}{ L_0 \lambda} \delta_{\alpha0}} 
e^{ -\pi l_\alpha^2 \frac{ L^2}{ L_\alpha^2 \lambda}}-1)  +r_0.
\ee

The expansion of $r_0$ follows immediately from \cite{HL}. The only 
modification is that the Goldstone boson mass in one of the terms now depends
on the chemical potential. We find
\be
r_0&=& -\log \frac{(m_\pi^2-4\mu^2)L^2}{4\pi} - \gamma -
\sum_{n=1}^\infty \frac 1{n!n} 
\left ( -\frac{(m_\pi^2-4\mu^2)L^2}{4\pi}\right )^n \nn \\
&&+\frac 12 \left (\frac{m_\pi^2 L^2}{4\pi}\right)^2
\left [\log\frac{m_\pi^2 L^2}{4\pi} + \gamma -\frac 32  \right ]
-\frac{1}{2}+ \frac{m_\pi^2 L^2}{4\pi}
+\sum_{n=3}^\infty \frac1{n!(n-2)}
\left (-\frac{m_\pi^2 L^2}{4\pi}\right )^n.
\ee
For the difference $g_0-r_0$ we obtain the expansion
\be
g_0 -r_0 = \sum_{n=0}^\infty \sum_{m=0}^\infty \frac 1{n! (2m)!}
\left [
\left (-\frac{m_\pi^2 L^2}{4\pi}\right )^n (4\mu^2 L^2)^m S_{n-2, m} 
+
\left (-\frac{(m_\pi^2-4\mu^2) L^2}{4\pi}\right )^n (-4\mu^2 L^2)^m 
S_{-n-2m, m} \right ]. 
\ee
For a $d$-dimensional hypercubic box 
the shape coefficients $S_{k,l}$ are given by
\be
S_{k,l>0} &=& \int_0^1 \frac {d\lambda}{\lambda} \lambda^k 
\sum_{l_0}l_0^{2l} e^{-\pi l_0^2/\lambda}
(\sum_{l_1} e^{-\pi l_1^2/\lambda})^{d-1},\nn\\
S_{k,0} &=& \int_0^1 \frac {d\lambda}{\lambda} \lambda^k 
[(\sum_{l_1} e^{-\pi l_1^2/\lambda})^{d} -1 ].
\ee
The shape coefficients $S_{k,0}$ already enter in the expansion for
$\mu=0$ \cite{HL}.  In the table below we give the numerical value of
a few low order coefficients.
\begin{table}[!h]
\begin{tabular}{c|c c c}
 $S_{k,l}$     & $l =0$  &$l =1$  &$l =2$  \\ 
\hline 
$k=0$  & \quad 0.09375685 & \quad 0.025131706 & \quad 0.025139306 \\
$k=-1$ & \quad 0.11745759 & \quad 0.031290174 & \quad 0.031298318\\
$k=-2$ & \quad  0.15365038 & \quad 0.040637153 & \quad 0.040645920\\
$k=-3$ & \quad  0.21251023 & \quad 0.055744392 & \quad 0.055753879\\
$k=-4$ & \quad  0.31550608 & \quad 0.082023256 & \quad 0.082033581\\
\end{tabular}
\end{table}
The lowest order terms of the expansion are given by
\be
g_0 &=&-\log \frac{(m_\pi^2-4\mu^2)L^2}{4\pi} - \gamma
-\frac{1}{2}+S_{0,0}+S_{-2,0} 
+\frac 12 \left ( \frac{m_\pi^2 L^2}{4\pi}\right )^2
\left [\log\frac{m_\pi^2 L^2}{4\pi} + \gamma -\frac 32  \right ] \nn \\
&&+\frac{(m_\pi^2-4\mu^2)L^2}{4\pi}(1-S_{-1,0}) 
+\frac{m_\pi^2 L^2}{4\pi}(1 -S_{-1,0}) \nn\\
&&+ \left( \frac{(m_\pi^2-4\mu^2)L^2}{4\pi} \right)^2(-\frac 14 +\frac 12 
S_{-2,0}) +\left (\frac{m_\pi^2 L^2}{4\pi} \right)^2  \frac 12 S_{0,0}
\nn \\
&&+ \frac{m_\pi^2 L^2}{8\pi} 4\mu^2L^2(-S_{-1,1} +S_{-3,1}) 
+ ( 4\mu^2L^2)^2(\frac{1}{24}S_{-2,2} +\frac{1}{24}S_{-4,2}- \frac 1{8\pi}
S_{-3,1}).  
\ee

\vspace{4mm}

\begin{figure}[!t]
  \unitlength1.0cm
  \epsfig{file=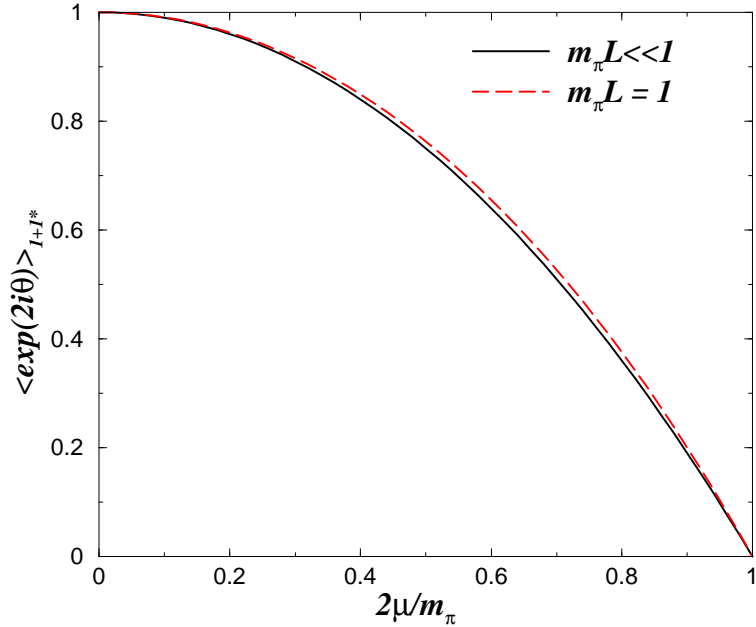,clip=,width=10cm}
  \caption{ \label{fig:1loop} The average phase factor in the phase quenched
    theory as a function of the chemical potential in the normal phase. The
    full line gives the result at leading order (\ref{exp-termo-n=1}) and the
    dashed line includes the corrections up to next-to-next-to-leading order
    (\ref{exp2ith-NNL}) when $m_\pi L=1$.}
\end{figure}

The $\mu$ independent terms do not contribute to the average phase
so that at next-to-next-to-leading order we find
\be
\langle e^{2i\theta}\rangle_{1+1^*} &=& (1-\frac{4\mu^2}{m_\pi^2})
\exp\left(\frac{4\mu^2L^2}{4\pi}(1-S_{-1,0})
    +\frac{8m_\pi^2\mu^2L^4-16\mu^4L^4}{(4\pi)^2}(-\frac 14 +\frac 12 
S_{-2,0})\right. \label{exp2ith-NNL} \\
&&\left.- \frac{m_\pi^2 L^2}{8\pi} 4\mu^2L^2(-S_{-1,1} +S_{-3,1}) 
- ( 4\mu^2L^2)^2(\frac{1}{24}S_{-2,2} +\frac{1}{24}S_{-4,2}- \frac 1{8\pi}
S_{-3,1})\right) .\nn
\ee
The $\mu L \to 0$, $m_\pi L \to 0$ limit of this expression should reproduce
the $\hat \mu\to \infty $, $\hat m \to \infty$ limit of the microscopic
result which is indeed the case (see (\ref{exp-termo-n=1})). The  
correction factor computed in (\ref{exp2ith-NNL}) drives the average 
phase factor closer to 1 (see figure \ref{fig:1loop}).   
Note that the correction is small even though $m_\pi L$ is set to 1.

\section {Conclusions}

We have computed the average phase factor in the microscopic domain 
of quenched, unquenched as well as phase quenched QCD and have found
that in the thermodynamic limit the average phase factor undergoes
a phase transition at $\mu = m_\pi/2$. 
For $\mu < m_\pi/2$
it falls off polynomially with $\mu$ and is equal to zero starting from
$\mu=m_\pi/2$. 
This implies that the phase problem becomes severe for $\mu > m_\pi/2$.
Physically this should be the case because the phase factor has to wipe out
the Bose condensate that would be the vacuum state if the fermion
determinant would have been  replaced by its absolute value. 

The results for the average phase factor in the microscopic limit of 
QCD were derived from a
chiral random matrix model exploiting the equivalence of these two theories
in the microscopic domain. Powerful random matrix methods such as
the method of orthogonal polynomials and
their Cauchy transforms enabled us to obtain exact analytical results.
The thermodynamic limit of these exact results,
was also derived from simple mean field arguments.
The starting point was that the average phase factor is 
the inverse ratio of the QCD partition function and the partition
function with the fermion determinant of QCD and an additional
conjugate bosonic quark and a fermionic quark.
For $\mu<m_\pi/2$ the 
partition function with the additional quarks is in the same phase as
the partition function without additional quarks. The ratio of the two
partition functions is therefore determined by the $1/V$ corrections to the 
free energies. For larger values of $\mu$ the presence of the conjugate
bosonic quark induces a pion condensate and the two  
free energies no longer cancel in the thermodynamic limit. 
This causes the exponential suppression with
the volume of the average phase factor.       

While the average phase factor is exponentially small for $\mu>m_\pi/2$ 
the fluctuations of the phase factor becomes unity for $\mu>m_\pi/2$. 
The sign problem is therefore particularly severe for $\mu>m_\pi/2$.

The finiteness of the average phase factor for $\mu<m_\pi/2$ suggests 
that it should be possible to test the predictions of this paper
by means of lattice QCD simulations. 
We note however that the quenched and unquenched phase
factor is non-analytic in $\mu$ at $\mu =0$. 
As has been shown explicitly in the quenched
case, the phase factor as defined by the ratio of two determinants with
opposite sign of the chemical potential, cannot be analytically continued
from negative $\mu^2$ to positive $\mu^2$. It is our experience that 
analyticity at $\mu=0$ is lost for averages involving the 
inverse (determinant) of the nonhermitian Dirac operator when the singularity
is not compensated by a similar factor from the fermion determinant.
In such cases lattice methods that rely on analyticity cannot be used.
On the other hand, we expect observables that are derivatives of the 
usual fermionic partition functions 
to be are analytic in $\mu$ for $ \mu \to 0$ and can
be computed by means of 
the Taylor expansion method or the imaginary chemical potential method.

Since each eigenvalue contributes to the phase factor it is perhaps
surprising that the average phase factor is not sensitive 
to the ultra-violet cutoff. The reason goes back to the renormalizablity
of chiral perturbation theory: If the theory is renormalized at $\mu =0$,
the $\mu$ dependence does not introduce any additional infinities.
Since the average phase factor is a partition function at $\mu\neq0$ 
divided by the same partition function $\mu =0$,
it is therefore necessarily ultraviolet finite.
We have shown this  explicitly to 1-loop order
for the phase quenched partition function.
This follows from the microscopic theory as well.
In lattice QCD the 
 Hasenfratz-Karsch prescription \cite{HK} 
to introduce the chemical potential does not give rise to additional
ultraviolet divergences in lattice perturbation theory.
However,  nonperturbatively the situation on the lattice 
is less clear.  Potentially dangerous ultraviolet
contributions may appear for a naive
evaluation of the phase factor of the fermion determinant on the lattice.
Despite of these and other potential pitfalls,
it is our hope that the results of this paper can be compared to  
lattice QCD and contribute constructively 
to our understanding of QCD at finite density.

\vspace{4mm}

\noindent
{\sl Acknowledgments.} 
We wish to thank S. Hands, C. Allton, G. Akemann, P.H. Damgaard, 
P. de Forcrand, M. L\"uscher, D. Dietrich, J. Osborn
and L. Ravagli for valuable discussions. This work was
supported  by U.S. DOE Grant No. DE-FG-88ER40388 (JV), the 
Carlsberg Foundation (KS), the Villum Kann Rassmussen Foundation (JV) and
the Danish National Bank (JV).

\newpage

\appendix

\section{Parameterization independence of mean field results.}
\label{app:jmass}

In this appendix we illustrate that the mean field result including the 
1-loop fluctuations about the saddle point (see eq. (\ref{gen1})) does not
depend on the parametrization of the integration manifold. The nontrivial
example we study is the phase quenched partition function in the condensed 
phase, i.e. for 
$\mu > m_\pi/2$. Because of the presence of an exactly massless mode
in this phase, we include a pion source which explicitly breaks the 
$U(1)$ symmetry that is responsible for the massless mode.
This allows us to use the saddle point
approximation to show that the partition function is independent of the
representation whereas the
Jacobian and masses are
representation dependent. 

The pion condensate source term enters in the mass matrix  
\be
\mat m & 0 \\ 0 & m \emat \to \mat m& \lambda \\ -\lambda & m \emat
\ee
so that the otherwise massless mode obtains a mass $\sim \sqrt \lambda$. 
We will consider two different representations of the Goldstone fields
\be
U = \mat \cos\theta e^{i\alpha} & e^{i\phi}\sin \theta\\
-e^{-i\phi}\sin \theta & \cos \theta e^{-i\alpha} \emat e^{i\omega}
\qquad {\rm and} \qquad U = e^{i\bar\theta\tau_1} e^{i\pi_k \tau_k}e^{i\omega},
\ee
where $\bar\theta$ is a vacuum angle that will be determined by the
saddle point equations. The Pauli matrices are labeled $\tau_k$.
The  partition function in either variables is given by
($\hat\lambda=\lambda\Sigma V$)
\be
Z_{1+1^*}^{\rm A}(\hat{m},\hat\lambda;\hat\mu) = \int d\theta d\alpha d\phi d\omega
\sin \theta \cos \theta 
e^{2\hat\mu^2 \sin^2 \theta + 
2\hat m \cos\theta \cos\alpha \cos\omega+ 2 \hat \lambda \sin\theta \cos\phi\cos \omega},
\label{ztheta}
\ee
and
\be
Z_{1+1^*}^{\rm B}(\hat{m},\hat{\lambda};\hat\mu) &=& \int d\pi_1 d\pi_2 d\pi_3 
d\omega \frac {\sin^2\pi}{\pi^2}
e^{\hat \mu^2[(\cos \bar\theta \frac {\sin\pi} \pi \pi_1 +\cos\pi\sin \bar\theta)^2+
(\cos \bar\theta \frac {\sin\pi} \pi \pi_2 
+\sin \bar\theta\frac {\sin\pi}\pi \pi_3)^2]}
\nn \\
&& \ \ \ \ \ \times e^{2\hat m[\cos \bar\theta\cos\pi 
- \sin \bar\theta\frac{\sin \pi}\pi \pi_1]\cos\omega
+\hat\lambda[\cos \bar\theta \frac {\sin\pi} \pi \pi_1 
+\cos\pi\sin \bar\theta]\cos\omega}
\ee
where $\pi = \sqrt{\pi_1^2+\pi_2^2+\pi_3^2}$.
We will evaluate both partition functions in the thermodynamic limit,
$\hat{\mu}^2 \to \infty$, $\hat{m} \to \infty$ and $\hat{\lambda} \to
\infty$, where the integrals can be performed by a saddle point
approximation. The parameter $\bar \theta$ will
be chosen such that the saddle point of the $\pi_k$ variables is at zero.
It turns out that this is also the saddle point of the $\theta$-integral
in (\ref{ztheta}).
In terms of the $\theta-\alpha-\phi-\omega$ variables the  squared masses
are given by
\be
\hat m\cos\bar \theta, 
\quad \hat \lambda \sin\bar \theta, \quad 2\hat \mu^2\sin^2\bar\theta 
+\frac{\hat \lambda}{\sin\bar \theta}, 
\quad \hat m \cos \bar \theta +\hat \lambda\sin \bar\theta
\ee
with $\bar \theta$ determined by the saddle point equation
\be
2\hat{\mu}^2\cos\bar\theta\sin\bar\theta -\hat m\sin \bar \theta +\hat\lambda \cos \bar \theta =0.
\label{saddleAB}
\ee
The squared masses in terms 
$\pi_k-\omega$-variables can be rewritten by using the saddle point
equation which is also given by (\ref{saddleAB}). We find
\be
2\hat{\mu}^2+\frac{\hat \lambda}{\sin \bar \theta}= \frac {\hat m}
{\cos \bar \theta}, 
\quad \frac {\hat \lambda}{\sin \bar \theta} ,
\quad 2\hat \mu^2 \sin^2 \bar \theta + \frac{ \hat \lambda} {\sin \bar \theta},
\quad \hat m \cos \bar \theta +\hat \lambda\sin \bar\theta.
\ee
The product of the squared masses is different for the two parameterizations.
The contribution from the Jacobian in (\ref{ztheta}),
$\sin \bar \theta \cos \bar \theta$ exactly compensates
for this difference such that  the mean field partition function (\ref{gen1})
is identical in the two cases. In the limit of $\hat \lambda \ll \hat m$ and
$\hat \lambda \ll \hat \mu^2$ we obtain
\be
Z_{1+1^*}(\hat{m},\hat{\lambda};\hat\mu) =\frac{\pi^2}
{\hat m \sqrt{\hat \lambda} \sqrt{2\hat \mu^2 - \hat m^2 /2 \hat\mu^2}}
e^{2\hat\mu^2 + \hat m^2 /2\hat\mu^2},
\ee
in agreement with the general expression (\ref{gen1}). 

What we have learned from this example is that the chemical potential
dependence of the partition function originates both from both 
the masses and the
Jacobian with neither of them being representation independent.

\noindent

\section{Integrals appearing in Cauchy transforms}
\label{app:id}

In this appendix we evaluate some integrals that appear in the 
Cauchy transforms $X^{(k)}$ given in (\ref{X0}).

To simplify the integrals appearing in $X^{(k)}$ the following
integral is useful
\be
\int_0^\infty dt t  e^{-p^2t^2} [K_0(at)I_0(bt) + I_0(at) K_0(bt)]
=\frac 1{2p^2} e^{\frac{a^2+b^2}{4p^2}} K_0\left ( \frac {ab}{2p^2}\right).
\ee
This identity can be proved by considering the integral
\be
\int_0^\infty dt t  e^{-p^2t^2} [K_0(at)I_0(bt) + I_0(at) K_0(bt)]
=\lim_{\nu \to 0}\int_0^\infty dt t  e^{-p^2t^2} [K_\nu(at)I_\nu(bt) 
+ I_{-\nu}(at) K_{-\nu}(bt)],
\ee
and using the identity 
\be
K_\nu(x) =\frac \pi 2 \frac{I_{-\nu}(x) - I_\nu(x)}{\sin(\nu\pi)}
\ee
to replace the $K_{\pm \nu}$ function by a $I_{\pm \nu}$ functions.

By differentiation with respect to the parameters of this integral, we
can derive the following  useful identities:
\be
\int_0^\infty dt t^2  e^{-p^2 t^2} [K_0(at)I_1(bt) - I_0(at) K_1(bt)]
&=&\frac 1{4p^4} e^{\frac{a^2+b^2}{4p^2}}\left[ b K_0\left ( \frac {ab}{2p^2}\right )
-a K_1\left ( \frac {ab}{2p^2}\right)\right] , \\
\int_0^\infty dt t^3  e^{-p^2 t^2} [K_0(at)I_0(bt) + I_0(at) K_0(bt)]
&=&\frac 1{4p^4} e^{\frac{a^2+b^2}{4p^2}}\left [(2+\frac{a^2+b^2}{2p^2}) 
K_0\left ( \frac {ab}{2p^2}\right )
-\frac{ab}{p^2} K_1\left ( \frac {ab}{2p^2}\right)\right],\\
\int_0^\infty dt t^4  e^{-p^2t^2} [K_0(at)I_1(bt) - I_0(at) K_1(bt)] &=& \nn
\\
&& \hspace{-4cm} \frac 1{8p^6} e^{\frac{a^2+b^2}{4p^2}}\left[(4b+\frac{b(3a^2+b^2)}{2p^2}) 
K_0\left ( \frac {ab}{2p^2}\right)
-\left (2a+ \frac {a(a^2+3b^2)}{2p^2}\right)  K_1\left (\frac{ab}{2p^2}\right) \right].
\ee

\section{Asymptotic Expansion of Integrals Occurring in the $X^{(k)}$}
\label{app:asymp}

In this section we calculate the asymptotic expansion of the integrals
occurring in the $X^{(k)}$ given in eq. 
(\ref{X0}). We separately consider the integral
\be
S_{p,\nu}=\int_0^{\hat{m}}dx x^p \exp[-\frac{x^2}{8\hat\mu^2}] K_0\left 
(\frac {x\hat{m}}{4\hat\mu^2}\right )
I_\nu(x )
\label{sint}
\ee
in the normal phase and in the condensed phase. 

In the normal phase, for $ \hat{m}/4 <\hat\mu^2 < \hat{m}/2$, the 
saddle point at $\bar x= 4\hat\mu^2-\hat{m}$ of the integral (\ref{sint})
is inside the integration domain. For $\hat{m} \to \infty$ and $\hat\mu^2 \to
\infty$ the Bessel functions can be expanded to leading order resulting in 
\be
S_{p, \nu}^{\rm normal} = 4\sqrt{\frac \pi{2\hat{m}}} 
\hat\mu^2 (4\hat\mu^2-\hat{m})^{p-1}
e^{2\hat{\mu}^2 -\hat{m} + \frac{\hat{m}^2}{8\hat{\mu}^2}}.
\ee
For $\hat\mu^2< \hat{m}/4$ the leading saddle point contribution comes from the second 
integral in the $X^{(k)}$. We leave it up to the reader to show that 
the leading order saddle point approximation results in the same expression.

In the condensed phase the leading order expansion of $S_{p,\nu}$ cancels
in the expression for the average phase. Therefore we have to
include the subleading terms in its asymptotic expansion. Including
the expansion of the 
 Bessel functions to subleading order we obtain for $S_{p,\nu}$
\be
S_{p,\nu}^{\rm condensed} \sim \sqrt{\frac{\hat{\mu}^2}{\hat{m}}}
\int_0^{\hat{m}}
dx x^{p-1} e^{-x^2/8\hat\mu^2 -x\hat{m}/4\hat\mu^2 +x} 
[1- \frac{\hat\mu^2}{2x\hat{m}}-\frac{4\nu^2-1}8 \frac 1{x}].
\ee
Next put $x= \hat{m} -t$. Since the integral is dominated by the vicinity 
of $t=0$ we can safely extend the integration range to $\infty$,
\be
S_{p,\nu}^{\rm condensed} 
\sim  \sqrt{\frac{\hat\mu^2}{\hat{m}}}e^{\hat{m}-\frac{3\hat{m}^2}{8\hat\mu^2}}
\int_0^\infty 
dt (\hat{m}^{p-1} -(p-1)\hat{m}^{p-2}t) e^{-t(1-\frac{\hat{m}}{2\hat\mu^2})-
\frac {t^2}{8\hat\mu^2}}
[1-\frac{\hat\mu^2}{2\hat{m}^2}-\frac{4\nu^2-1}8 \frac 1{\hat{m}}].
\label{tint}
\ee
To leading nonvanishing order it suffices to expand
 $\exp(-t^2/8\hat\mu^2)$ as $ 1- \frac{t^2}{8\hat\mu^2}$. After 
performing the integral over $t$ we find
\be
S_{p,\nu}^{\rm condensed} 
=  \sqrt{\frac{\hat\mu^2}{\hat{m}}}e^{\hat{m}-\frac{3\hat{m}^2}{8\hat\mu^2}}
\hat{m}^{p-2}\left [\frac{2\hat{m}}{2-\frac{\hat{m}}{\hat\mu^2}} 
 -\frac {4(p-1)}{(2-\frac{\hat{m}}{\hat\mu^2})^2} - \frac{2\hat{m}}{\hat\mu^2(2-\hat{m}/\hat\mu^2)^3}
-\frac 1{(2-\hat{m}/\hat\mu^2)}[\frac{\hat\mu^2}{\hat{m}}+\frac{4\nu^2-1}{4} 
]\right ].\nn\\
\ee
Combining the different terms we obtain for the thermodynamic limit
of
the expectation value of the quenched phase
\be
\langle \exp(2i\theta) \rangle_{N_f=0} = \frac 1{\sqrt{2\pi \hat{m}}}
\frac {1}{\hat\mu^2(2-\hat{m}/\hat\mu^2)^2}\sqrt{\frac{\hat\mu^2}{\hat{m}}} 
e^{2\hat{m}-\hat{m}/2\hat\mu^2 -2\hat\mu^2}
\qquad {\rm for} \qquad 1\ll \hat m < 2\hat\mu^2.   
\label{c6}  
\ee
The condition for the validity of the derivation of (\ref{c6}) is that
$(2\hat \mu^2 -\hat m)/\hat \mu \gg 1$, which is only violated very
close to the critical point because the natural magnitude of this
ratio  of $O(\sqrt V)$. 
We can also evaluate the integral (\ref{tint}) when this condition 
 is not satisfied. 
Then the integral (\ref{tint}) is cut-off by the
the Gaussian factor $\exp(-t^2/2\hat\mu^2)$ instead of the exponential factor
resulting in an expression that remains finite for $m=2\mu^2$.

\end{document}